\documentclass[twocolumn]{aastex7}

\usepackage{makecell}
\usepackage{amsmath}
\usepackage[labelfont=bf]{caption}
\usepackage{array} % ensures newcolumntype is available

\begin{document}

\title{Demonstrating Exoplanet Transit Photometry from Space\\with a 15-mm Aperture Optical Navigation Camera on Hayabusa2}

\author[orcid=0000-0001-6160-9360,gname=Koki,sname=Yumoto]{Koki Yumoto}
\affiliation{LIRA, Observatoire de Paris, Université PSL, Sorbonne Université, Université Paris-Cité, CY Cergy Paris Université, CNRS, \newline Meudon, France}
\affiliation{Institute of Space and Astronautical Science, Japan Aerospace Exploration Agency, Sagamihara, Kanagawa, Japan}
\email[show]{kyyumoto@gmail.com}  

\author[gname=Toru, sname=Kouyama]{Toru Kouyama} 
\affiliation{Artificial Intelligence Research Center, National Institute of Advanced Industrial Science and Technology, Koto, Tokyo, Japan}
\email{t.kouyama@aist.go.jp}

\author[gname=Manabu, sname=Yamada]{Manabu Yamada} 
\affiliation{Planetary Exploration Research Center (PERC), Chiba Institute of Technology, Narashino, Chiba, Japan}
\email{manabu@perc.it-chiba.ac.jp}

\author[orcid=0000-0002-7181-3522,gname=Yuya, sname=Mimasu]{Yuya Mimasu} 
\affiliation{Institute of Space and Astronautical Science, Japan Aerospace Exploration Agency, Sagamihara, Kanagawa, Japan}
\email{mimasu.yuya@jaxa.jp}

\author[gname=Tomokatsu, sname=Morota]{Tomokatsu Morota} 
\affiliation{Department of Earth and Planetary Science, The University of Tokyo, Bunkyo, Tokyo, Japan}
\email{morota@eps.s.u-tokyo.ac.jp}

\author[orcid=0000-0003-2749-2204,gname=Yuichiro, sname=Cho]{Yuichiro Cho} 
\affiliation{Department of Earth and Planetary Science, The University of Tokyo, Bunkyo, Tokyo, Japan}
\email{cho@eps.s.u-tokyo.ac.jp}

\author[orcid=0000-0001-9230-974X,gname=Yasuhiro, sname=Yokota]{Yasuhiro Yokota} 
\affiliation{Institute of Space and Astronautical Science, Japan Aerospace Exploration Agency, Sagamihara, Kanagawa, Japan}
\affiliation{Institute of Science Tokyo, Ookayama, Tokyo, Japan}
\email{yokota@planeta.sci.isas.jaxa.jp}

\author[gname=Masahiko, sname=Hayakawa]{Masahiko Hayakawa} 
\affiliation{Institute of Space and Astronautical Science, Japan Aerospace Exploration Agency, Sagamihara, Kanagawa, Japan}
\email{hayakawa.masahiko@jaxa.jp}

\author[orcid=0000-0001-5097-4784,gname=Anthony, sname=Arfaux]{Anthony Arfaux} 
\affiliation{LIRA, Observatoire de Paris, Université PSL, Sorbonne Université, Université Paris-Cité, CY Cergy Paris Université, CNRS, \newline Meudon, France}
\email{anthony.arfaux@obspm.fr}

\author[gname=Eri, sname=Tatsumi]{Eri Tatsumi}
\affiliation{Institute of Space and Astronautical Science, Japan Aerospace Exploration Agency, Sagamihara, Kanagawa, Japan}
\affiliation{Instituto de Astrofísica de Canarias (IAC), University of La Laguna, La Laguna, Tenerife, Spain}
\affiliation{Department of Astrophysics, University of La Laguna, La Laguna, Tenerife, Spain}
\email{eri.tatsumi@iac.es}

\author[gname=Moe, sname=Matsuoka]{Moe Matsuoka}
\affiliation{Geological Survey of Japan (GSJ), National Institute of Advanced Industrial Science and Technology, Tsukuba, Ibaraki, Japan}
\email{moe.matsuoka@aist.go.jp}

\author[orcid=0000-0003-4791-5227,gname=Naoya, sname=Sakatani]{Naoya Sakatani}
\affiliation{Institute of Space and Astronautical Science, Japan Aerospace Exploration Agency, Sagamihara, Kanagawa, Japan}
\email{sakatani.naoya@jaxa.jp}

\author[gname=Sumito, sname=Shimomura]{Sumito Shimomura}
\affiliation{Institute of Space and Astronautical Science, Japan Aerospace Exploration Agency, Sagamihara, Kanagawa, Japan}
\email{shimomura.sumito@jaxa.jp}

\author[gname=Shingo, sname=Kameda]{Shingo Kameda}
\affiliation{Department of Physics, Rikkyo University, Toshima, Tokyo, Japan}
\email{kameda@rikkyo.ac.jp}

\author[gname=Satoshi, sname=Tanaka]{Satoshi Tanaka}
\affiliation{Institute of Space and Astronautical Science, Japan Aerospace Exploration Agency, Sagamihara, Kanagawa, Japan}
\email{tanaka.satoshi@jaxa.jp}

\author[gname=Keigo, sname=Enya]{Keigo Enya}
\affiliation{Institute of Space and Astronautical Science, Japan Aerospace Exploration Agency, Sagamihara, Kanagawa, Japan}
\email{enya.keigo@jaxa.jp}

\author[orcid=0000-0001-6076-3614,gname=Seiji, sname=Sugita]{Seiji Sugita}
\affiliation{Department of Earth and Planetary Science, The University of Tokyo, Bunkyo, Tokyo, Japan}
\affiliation{Planetary Exploration Research Center (PERC), Chiba Institute of Technology, Narashino, Chiba, Japan}
\email{sugita@eps.s.u-tokyo.ac.jp}

\begin{abstract}
Observations of exoplanet transits by small satellites have gained increasing attention for reducing detection biases. However, no unambiguous detection of an exoplanet has yet been demonstrated using optics with apertures smaller than 60 mm. Here, we investigated the detectability of exoplanet transits using the telescopic Optical Navigation Camera (ONC-T) onboard the Hayabusa2 spacecraft, which has an effective aperture of only 15 mm. We conducted transit observations of the hot Jupiters WASP-189 b and MASCARA-1 b, collecting data for ten and four events, respectively. The transit signal was detected with a signal-to-noise ratio (SNR) of 13 for WASP-189 b and 8 for MASCARA-1 b for each event. Stacking all events improved the SNR to 40 and 16, respectively. The transit mid-times of each event were measured with a precision of 6 minutes and were consistent with Transiting Exoplanet Survey Satellite (TESS) data to within 2 minutes. The planet-to-star radius ratio was determined with an absolute precision of 0.004 (6\% relative) and agreed with TESS results to within 0.002 (3\% relative). The recent ONC-T and TESS data enabled an update to the planetary ephemerides. We report a 4$\sigma$ discrepancy between the updated orbital period of MASCARA-1 b and previously reported values. ONC-T sets a new record for the smallest-aperture instrument to detect an exoplanet transit from space, advancing the frontier of exoplanet science with miniature instrumentation. Our results suggest that optics as small as ONC-T may be capable of detecting transiting long-period Jupiters: a population that remains underrepresented in current surveys.
\end{abstract}



\section{Introduction}\label{Intro}

\begin{figure}[ht]
    \centering
    \includegraphics[width=0.9\hsize]{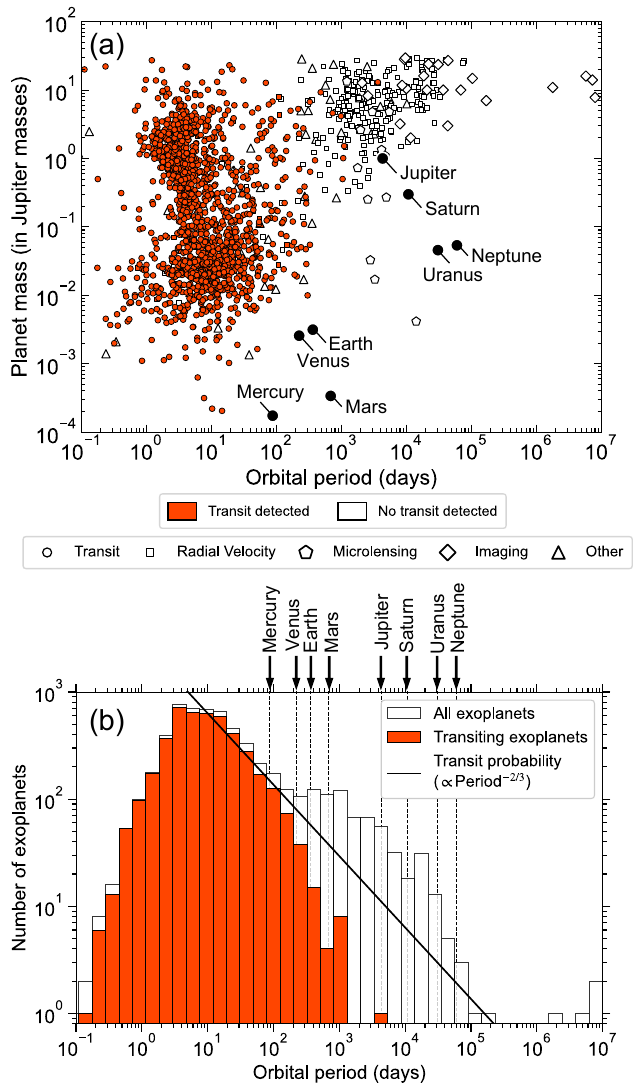}
    \caption{
    Population statistics of all 5,917 confirmed exoplanets as of 11 June 2025. \textbf{(a)} Orbital period vs. planet mass for all planets with direct mass estimates. Red filled symbols represent exoplanets with transit detections, while white open symbols denote those without. Marker shapes indicate the method of discovery. \textbf{(b)} Distribution of orbital periods for all known exoplanets and transiting exoplanets. The black solid line represents the transit probability, proportional to Period$^{-2/3}$, scaled to match the data at a 10-day period. This line represents the expected distribution of transiting exoplanets, under the assumption that planet occurrence is uniform in $\log_{10} (\mathrm{Period}$) space. Data are from the NASA Exoplanet Archive.
    }
    \label{Fig_exoplanets}
\end{figure}

Despite the discovery of over 5,000 exoplanets, those with orbits and masses similar to planets in our Solar System remain limited in number; most known exoplanets have larger masses or shorter orbital periods (Fig. \ref{Fig_exoplanets}a). This apparent absence may be largely driven by observational biases (e.g., \citealt{Foreman-Mackey_2016}). Radial velocity (RV) surveys are capable of detecting planets with year-long orbits, since they do not necessitate dense temporal sampling across the full orbital period. However, their sensitivity is generally limited to planets more massive than approximately one Jupiter mass, due to the weak RV signals exerted by lower-mass planets (\citealt{Ananyeva_2022}). In contrast, transit observations can detect planets across a broader range of sizes, but the current sample likely lacks long-period planets because their detection requires continuous, long-duration time-series data. The number of transiting planets identified to date begins to fall short of that expected for a uniform distribution of exoplanets in $\log_{10}(\mathrm{Period})$ space once orbital periods exceed $\sim$100 days (Fig. \ref{Fig_exoplanets}b). While this apparent deficit could reflect a genuine scarcity of long-period planets, it is more due to observational bias, as RV surveys find a nearly uniform distribution over the $\log_{10}(\mathrm{Period})$ domain \citep{Cumming_2008}. For instance, the \textit{Transiting Exoplanet Survey Satellite} (TESS) observes a given field for only $\sim$27 days, limiting its ability to detect longer-period planets. Mitigating these biases through additional surveys is essential to assess whether Solar System-like planets are unique and to better understand the formation pathways of habitable planets (\citealt{NationalAcademies_2018}).

% -------------------------------------------- %
% ------------- Begin Table 1 ---------------- %
% -------------------------------------------- %

\begin{deluxetable*}{lllllll}
\setlength\tabcolsep{1.3pt} % column spacing
\renewcommand{\arraystretch}{0.7} % row spacing
\tabletypesize{\scriptsize}
\tablecaption{Performance metrics of miniature optics applied to exoplanet transit observations.\label{Table_review}}

\tablehead{
\colhead{Instrument} & 
\colhead{Wavelength} & 
\colhead{Plate scale} & 
\colhead{Aperture} & 
\colhead{\parbox[c]{3.2cm}{\centering Detector \\ temperature stability\\[0.9ex]}} & 
\colhead{Pointing stability} & 
\colhead{Transit detections\rule{0pt}{5.0ex}}
}
\startdata
\rule{0pt}{3.0ex}
Hayabusa2/& 400--950 nm$^{(1)}$ & 22.1" pixel$^{-1}$$^{(1)}$ & 15 mm$^{(1)}$ & 2--4$^\circ\!$C (RMSD)$^{(2)}$ & 6–17" (RMSD)$^{(2)}$ & Detected transits of\\
\hspace{1ex}ONC-T&&&&&&two known hot Jupiters$^{(2)}$.\\[1.0ex]
\hline
\rule{0pt}{3.0ex}
STEREO/& 630–730 nm$^{(3)}$ & 35" pixel$^{-1}$$^{(3)}$ & 16 mm$^{(3)}$ & Passively cooled to & 7$''$ (RMSD)$^{(3)}$ & Found transit candidates \\
\hspace{1ex}HI-1A&&&&$\sim$ –80$^\circ\!$C$^{(4)}$&&(follow-up required)$^{(3,5)}$.\\[1.0ex]
\hline
\rule{0pt}{3.0ex}
BRITE & 400–450 nm or & 27–30" pixel$^{-1}$$^{(6)}$ & 3 cm$^{(6)}$ & $\sim$4$^\circ\!$C drift & 45" (RMSD)$^{(8)}$ & Found transit candidates\\
\hspace{1ex}constellation&550–700 nm$^{(6)}$ &&&per orbit$^{(7,a)}$&&(follow-up required)$^{(9,10)}$.\\
&&&&&&No detection in the \\
&&&&&&$\beta$ Pictoris system$^{(11)}$.\\[1.0ex]
\hline
\rule{0pt}{3.0ex}
PicSat & 400–680 nm$^{(12)}$ & 4.2" pixel$^{-1}$$^{(13,b)}$ & 3.7 cm$^{(12)}$ & 0.004$^\circ\!$C (RMSD)$^{(14)}$ & 30" (RMSD)$^{(12,c)}$ & No data due to\\
&&&&&&attitude failure$^{(14)}$.\\[1.0ex]
\hline
\rule{0pt}{3.0ex}
ASTERIA & 500–900 nm$^{(15)}$ & 15.8" pixel$^{-1}$$^{(15)}$ & 6.07 cm$^{(15)}$ & $\pm$0.01$^\circ\!$C & 0.5" (RMSD)$^{(15)}$ & Marginal detection of \\
&&&&(max–min)$^{(15)}$ &&55 Cancri e$^{(16)}$. \\
&&&&&&No detection in the $\alpha$ Centauri\\
&&&&&&A and B systems$^{(17)}$.\\[1.0ex]
\hline
\rule{0pt}{3.0ex}
TESS & 600–1000 nm$^{(18)}$ & 21" pixel$^{-1}$$^{(18)}$ & 10.5 cm$^{(18)}$ & $<$0.02$^\circ\!$C per day$^{(19)}$ & $<$2.0" (3$\sigma$)$^{(20)}$ & Detected $>$7000 transits\\
&&&&&&including candidates$^{(\textup{e.g.,}\ 21)}$.\\[1.0ex]
\hline
\rule{0pt}{3.0ex}
CUTE & 247.9–330.6 nm & 2.5" pixel$^{-1}$$^{(22)}$ & 20.6$\, \times\, $8.4 cm$^{(22)}$ & $\pm$2$^\circ\!$C (max–min)$^{(23)}$ & $<$6" (RMSD)$^{(23)}$ & Detected the transit spectrum\\
&(0.39 nm steps)$^{(22)}$ &&&&&of WASP-189 b$^{(24)}$.\\[1.0ex]
\hline
\rule{0pt}{3.0ex}
MOST & 380–700 nm$^{(25)}$ & 3" pixel$^{-1}$$^{(25)}$ & 15 cm$^{(25)}$ & $\pm$0.1$^\circ\!$C$^{(25)}$ & 3–5" (RMSD)$^{(26)}$ & Detected transits in \\
&&&&&&multiple systems$^{(\textup{e.g.,}\ 27,28)}$.\\[1.0ex]
\hline
\enddata
\tablecomments{(a) The magnitude of the drift varies depending on the satellite’s orientation relative to the Sun.\\
(b) Defined by the 3\,µm diameter of the fiber core.\\
(c) Additional photometric stability is achieved through fine adjustments of the fiber head position within the focal plane.\\
\textbf{References:}
(1) \citet{Kameda_2017};
(2) This study;
(3) \citet{Wraight_2011};
(4) \citet{Eyles_2009};
(5) \citet{Whittaker_2014};
(6) \citet{Weiss_2021};
(7) \citet{Popowicz_2018};
(8) \citet{Sarda_2014};
(9) \citet{Yeh_2020};
(10) \citet{Fuentes_2024};
(11) \citet{Lous_2018};
(12) \citet{Lapeyrere_2017};
(13) \citet{Nowak_2017};
(14) \citet{Nowak_2018};
(15) \citet{Smith_2018};
(16) \citet{Knapp_2020};
(17) \citet{Krishnamurthy_2021};
(18) \citet{Ricker_2015};
(19) \citet{Vanderspek_2018};
(20) \citet{Nguyen_2018};
(21) \citet{Ivshina_2022};
(22) \citet{France_2023};
(23) \citet{Egan_2023};
(24) \citet{Sreejith_2023};
(25) \citet{Walker_2003};
(26) \citet{Grocott_2004};
(27) \citet{Winn_2011};
(28) \citet{Miller-Ricci_2008}.
}                                                               
\end{deluxetable*}

% -------------------------------------------- %
% --------------- End Table 1 ---------------- %
% -------------------------------------------- %

One proposed strategy for reducing these biases is to search for transit signals around bright stars through long-term monitoring with small satellites (e.g., \citealt{Kawahara_2020}). The use of small satellites for transit detection has attracted attention owing to their stable, atmosphere-independent performance and the substantial recent reduction in launch costs (\citealt{Shkolnik_2018}; \citealt{Serjeant_2020}). These advantages enable dedicated, long-term monitoring of individual stars (\citealt{Millan_2019}), and many key technologies have been demonstrated in recent missions. For instance, the 6U CubeSat \textit{Arcsecond Space Telescope Enabling Research in Astrophysics} (ASTERIA) successfully demonstrated the capability of small satellites to maintain long-term pointing and thermal stability (Table \ref{Table_review}; \citealt{Smith_2018}). The \textit{Bright Target Explorer} (BRITE) constellation, consisting of 20-cm nanosatellites, has demonstrated that deploying a network of nearly identical satellites can significantly improve both temporal and spatial coverage (\citealt{Weiss_2021}). Building on these demonstrations, the \textit{LOng-period Transiting exoplanet sUrvey Satellite} (LOTUS) being developed at the Japan Aerospace Exploration Agency (JAXA) plans to use a nanosatellite equipped with a 7.5-cm aperture wide-field imager to search for long-period planets that remain undetected (\citealt{Kawahara_2020}).

Despite these promising advances in small-satellite technologies, their photometric capability for transit detection has yet to be fully explored. To date, the marginal 2.2$\sigma$ detection of the super-Earth 55 Cnc e by ASTERIA, equipped with a 60 mm aperture optics, remains the most compact optical system to yield a credible transit signal, while no unambiguous detection has been reported with smaller aperture optics (Table \ref{Table_review}). Several transit-like signals have been identified with the 30-mm aperture of BRITE (\citealt{Yeh_2020, Fuentes_2024}) and the 16-mm aperture of \textit{Solar Terrestrial Relations Observatory}/\textit{Heliospheric Imager-1A} (STEREO/HI-1A; \citealt{Wraight_2011}; \citealt{Whittaker_2014}), but it remains uncertain whether these signals originate from actual exoplanet transits. For instance, the transit-like signal of HD 213597 detected by STEREO/HI-1A was later found to be an eclipsing binary (\citealt{Chaturvedi_2014}), while other candidates lack sufficient follow-up data for confirmation. The lack of unambiguous transit detections by BRITE and STEREO/HI-1A does not necessarily indicate limitations of small-aperture optics, as their observations were neither targeted at nor timed to capture well-characterized exoplanets. Thus, optics with apertures smaller than 60 mm may still be viable for transit detection, and their sensitivity to transit signals needs to be quantified to assess the capabilities of future small-satellite missions.

To address this issue, we conducted transit observations of two known hot Jupiters using the \textit{telescopic Optical Navigation Camera} (ONC-T) onboard the Hayabusa2 spacecraft: an asteroid sample return mission led by JAXA. With an effective aperture of just 15 mm, ONC-T represents, to our knowledge, the smallest aperture ever used for transit observations from space (Table \ref{Table_review}). Its pointing and thermal stability are comparable to those demonstrated by previous small satellite missions such as BRITE and CUTE, with ASTERIA exhibiting even greater stability (Table \ref{Table_review}). In addition, ONC-T exhibits a high number of hot pixels resulting from its 10 years of operation (\citealt{Kouyama_2021}), simulating the degraded detector conditions that future missions may experience after long-term operations. These features make ONC-T a unique space-based photometer for studying the scientific potential of compact optics deployable on CubeSats and nanosatellites. In this study, we assess the detectability of transits by ONC-T and evaluate the measurement accuracy of transit depths and timings by comparing the results with those from TESS and previous observations.

% --------------------------------------------- %
% --------------- Observations ---------------- %
% --------------------------------------------- %

\section{Observations}
\label{Obs}

\subsection{Telescopic Optical Navigation Camera (ONC-T) onboard Hayabusa2}
\label{Obs:ONC}

ONC-T is a multi-band, shutter-less charge-coupled device (CCD) imager used for both scientific observations and onboard navigation. It played a crucial role in guiding Hayabusa2 to asteroid (162173) Ryugu and in precisely targeting the sample collection sites (\citealt{Ogawa_2020}). The dioptric system of ONC-T has a focal length of 120.5 mm (\citealt{Suzuki_2018}) and an effective aperture diameter of 15 mm (\citealt{Kameda_2017}). The camera is equipped with a filter wheel containing one panchromatic filter covering the 400–950 nm wavelength range and seven narrow-band filters. The detector is CCD4720AIMO (Teledyne e2v), which has a 1024 $\times$ 1024-pixel array with a pixel size of 13 µm and a 100\% fill factor, i.e., the entire detector surface is sensitive to light. The detector is passively cooled to –30°C during nominal observations, with temperature stability maintained via an onboard heater. Although ONC-T has operated nominally since launch, the number of hot pixels has increased linearly over time, attributed to cumulative radiation damage to the CCD (\citealt{Kouyama_2021}).

\begin{deluxetable*}{cccccc}
\setlength\tabcolsep{2.4pt} % column spacing
\renewcommand{\arraystretch}{1} % row spacing
\tabletypesize{\scriptsize}
\tablecaption{List of transit events observed by the Optical Navigation Camera along with their observational conditions.\label{Table_obscond}}

\tablehead{
\colhead{Event} & 
\colhead{Planet name} & 
\colhead{Observation time (UTC)} & 
\colhead{\parbox[c]{2.2cm}{\centering Imaging \\ cadence (sec)\\[0.9ex]}} & 
\colhead{\parbox[c]{2.2cm}{\centering Exposure time \\ per image (sec)\\[0.9ex]}} & 
\colhead{\parbox[c]{2.2cm}{\centering Number \\ of images\\[0.9ex]}\rule{0pt}{5.0ex}}
}
\startdata
    1 & WASP-189 b & 2022 Dec 17 19:00:27 – 2022 Dec 18 18:58:27 & 120 & 44.56 & 720\\
    2 & WASP-189 b & \hspace{3.5ex}2022 Dec 25 18:30:28 – 2022 Dec 26 18:28:29 $^{(a)}$& 120 & 44.56 & 702\\
    3 & WASP-189 b & 2023 Jan 05 18:40:24 – 2023 Jan 06 18:38:25 & 120 & 44.56 & 720\\
    4 & WASP-189 b & 2023 Jan 08 12:03:24 – 2023 Jan 09 12:01:24  & 120 & 44.56 & 720\\
    5 & WASP-189 b & 2023 Jan 13 22:48:26 – 2023 Jan 14 19:46:25 & 120 & 44.56 & 630\\
    6 & WASP-189 b & 2023 Jan 19 09:34:24 – 2023 Jan 20 06:32:25 & 120 & 44.56 & 630\\
    7 & WASP-189 b & 2023 Jan 22 02:56:25 – 2023 Jan 22 23:54:25 & 120 & 44.56 & 630\\
    8 & WASP-189 b & 2023 Nov 12 09:00:28 – 2023 Nov 13 06:13:23 & 65 & 44.56 & 1176\\
    9 & WASP-189 b & 2024 Sep 15 04:00:26 – 2024 Sep 16 01:13:21 & 65 & 44.56 & 1176\\
    10 & WASP-189 b & 2024 Sep 17 21:00:24 – 2024 Sep 18 18:13:19 & 65 & 44.56 & 1176\\
    11 & MASCARA-1 b & 2022 Jun 11 10:01:31 – 2022 Jun 12 09:57:31 & 240 & 178.26 & 360\\
    12 & MASCARA-1 b & 2024 Nov 29 23:01:32 – 2024 Nov 30 20:18:43 & 200 & 178.26 & 384\\
    13 & MASCARA-1 b & 2024 Dec 04 04:01:31 – 2024 Dec 05 01:18:42 & 200 & 178.26 & 384\\
    14 & MASCARA-1 b & 2024 Dec 21 09:01:31 – 2024 Dec 22 06:18:42 & 200 & 178.26 & 384\\
    \hline
\enddata

    \tablecomments{(a) Data gap between 22:06:29–22:14:29 and 22:52:29–23:16:29 on 2022 Dec 25.}
    
\end{deluxetable*}

Building on the high photometric precision of ONC-T demonstrated over years of operation \citep{Yumoto_2024, Tatsumi_2019}, we launched a campaign to observe exoplanet transits during the extended mission’s 10-year space cruise (2021–2031). We focused on observing confirmed exoplanets, where transit timings are predictable. The targets were selected based on their visibility and the expected signal-to-noise ratio (SNR). Hayabusa2 orbits the Sun at approximately 1 AU in the ecliptic plane, completing one revolution per year. In its nominal configuration, the spacecraft is oriented such that the ONC-T boresight points approximately in the anti-solar direction, with a $\sim$10° offset to optimize the influence of solar radiation pressure on spacecraft dynamics (\citealt{Tsuda_2017}). Due to constraints related to spacecraft thermal stability, solar radiation pressure, and stray lights, ONC-T can point within a favorable angular range of ±10°, with a hard limit of ±40°, from the nominal pointing direction. As a result, stars located within ±10° or ±40° of the ecliptic plane are observable, each with an annual visibility window of approximately two weeks to one month. For all confirmed transiting exoplanets within this range, we calculated the predicted SNR of their transit signals (see Section \ref{Discussion:Implications} for details on the calculation). 

WASP-189 b and MASCARA-1 b were selected based on their high predicted SNR, ranking 2nd and 9th, respectively, among all visible targets. Both planets are ultra-hot Jupiters with $\gtrsim$2 Jupiter masses and dayside temperatures exceeding 3000 K. WASP-189 b orbits a $m_{V}=$ 6.6 star with a 2.7-day period (\citealt{Lendl_2020}), and MASCARA-1 b orbits a $m_{V}=$ 8.3 star with a 2.1-day period (\citealt{Hooton_2022}). These planets also warrant a search for potential transit timing variations (TTVs), as their ephemerides were not updated in the most recent TTV surveys by \citet{Ivshina_2022} and \citet{Kokori_2023}, which included data only through 2021 due to the limited availability of recent observations at the time.

We observed 10 transits of WASP-189 b and four transits of MASCARA-1 b over a period of 2.5 years. The observation times and imaging conditions are summarized in Table \ref{Table_obscond}. Each transit was observed continuously for approximately 21 hours, covering at least $\pm$8 hours around the predicted mid-transit time and fully capturing the ingress and egress. The transits were observed using the panchromatic filter to maximize the SNR. For each observation, the spacecraft was oriented to center the target star within the recorded portion of the image and to minimize stray light (\citealt{Tatsumi_2019}). The spacecraft attitude was held fixed throughout each observation. The duration of each observation and the interval between successive sessions were constrained by the buildup of solar radiation pressure, caused by the non-optimal spacecraft orientation required during observations; minimizing this load has been essential to maintain attitude control without fuel consumption (\citealt{Tsuda_2017}). The imaging cadence ranged from 65 to 240 seconds, including both the exposure time and the overhead needed to store each image in the onboard data recorder.

\begin{figure*}[th]
    \centering
    \includegraphics[width=0.72\hsize]{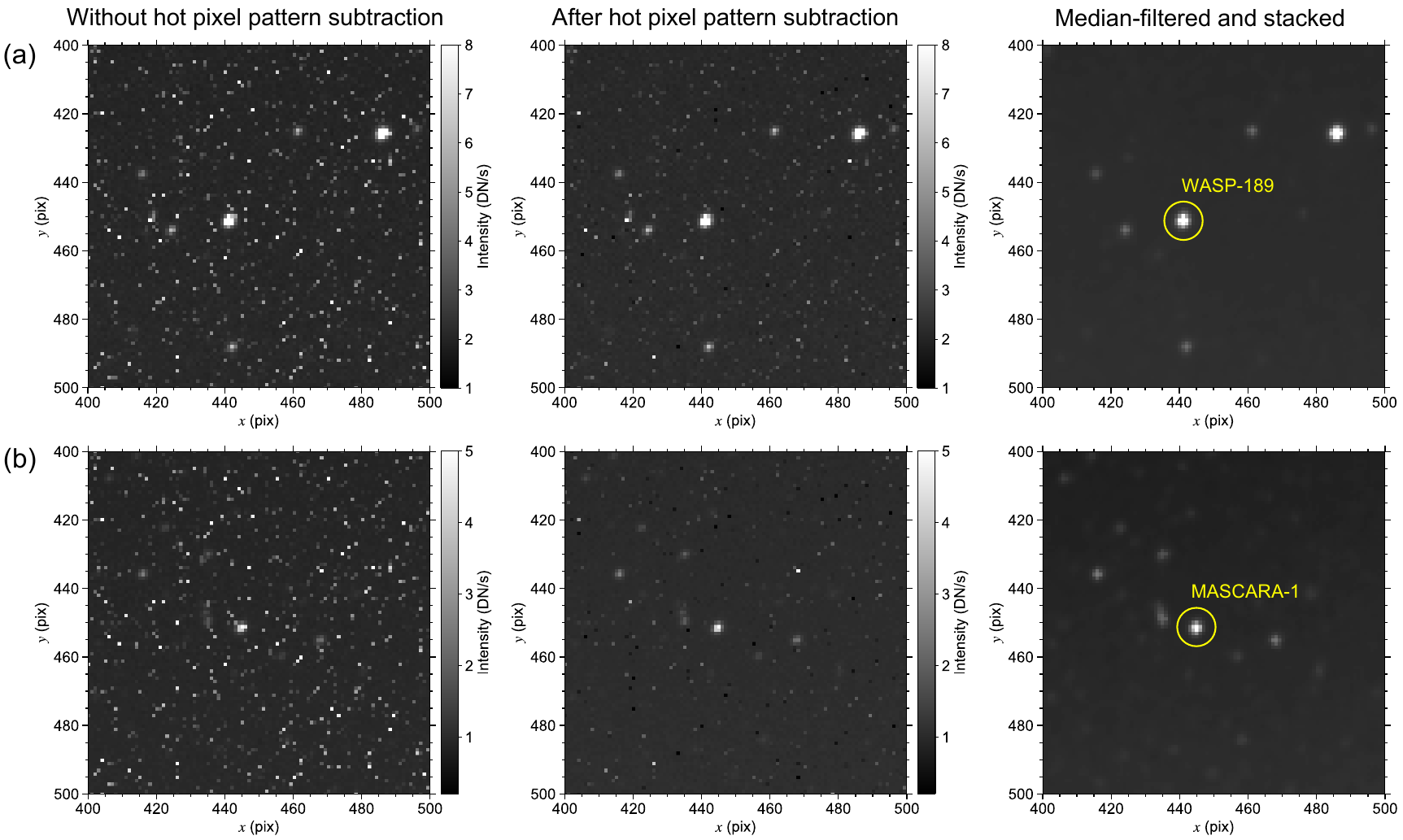}
    \caption{
    ONC-T images of \textbf{(a)} WASP-189 (image hyb2\_onc\_20240918\_073620\_tif\_l2b) and \textbf{(b)} MASCARA-1 (image hyb2\_onc\_20241130\_093826\_tif\_l2b) captured during transit events 10 and 12, respectively. The left panels show a single reduced ONC-T image ($p^t (x,y)$) without hot pixel correction (i.e., data reduced assuming $p_\mathrm{hot}=0$ in Equation \ref{Eq1}). The center panels show the same reduced images with hot pixel subtraction. The right panels show the reduced ONC-T image after applying a 3 $\times$ 3-pixel median filter and stacking all frames co-aligned on the target star, allowing the stellar positions to be clearly visualized. The host stars are marked with circles.
    }
    \label{Fig_oncim}
\end{figure*}

\subsection{Transiting Exoplanet Survey Satellite (TESS)}
\label{Obs:TESS}

We analyzed TESS data for WASP-189 b and MASCARA-1 b, acquired within eight months of the ONC-T observations, to validate the ONC-T results. WASP-189 b was observed by TESS in Sector 51 (April 22–May 18, 2022), while MASCARA-1 b was observed in Sectors 55 (August 5–September 1, 2022) and 82 (August 10–September 5, 2024). Transit events without continuous coverage from ingress to egress were excluded, resulting in three usable transits for WASP-189 b and 20 for MASCARA-1 b (9 in Sector 55 and 11 in Sector 82).

% --------------------------------------------- %
% ------- Data reduction and modeling --------- %
% --------------------------------------------- %

\section{Data reduction and modeling}
\label{DataRedModel}

\subsection{Light curves from aperture photometry}
\label{DataRedModel:LC}

The raw ONC-T images were first corrected for bias, dark current, and hot pixels, then divided by the flat-field pattern and normalized by the exposure time:
\begin{multline}
    \label{Eq1}
    p^{t}(x,y) =
    \Bigl[
      p^{t}_{\mathrm{raw\,}}(x,y)
      - p^{t}_{\mathrm{bias\,}}\!(T_{\mathrm{CCD}}^{t}, T_{\mathrm{ELE}}^{t}, T_{\mathrm{AE}}^{t}) \\
      - p^{t}_{\mathrm{dark\,}}\!(T_{\mathrm{CCD}}^{t})
      - p_{\mathrm{hot\,}}\!\left(x,y,T_{\mathrm{CCD}}\right)
    \Bigr]\cdot\frac{1}{t_{\mathrm{exp}}\, f(x,y)}.
\end{multline}
Here, $p^{t}_{\mathrm{raw}}(x,y)$ represents the raw signal intensity (in DN units) of an image taken at time $t$ and pixel coordinates $(x,y)$, while $p^{t}(x,y)$ denotes the corresponding image after reduction. The exposure time is given by $t_\mathrm{exp}$ in seconds, and $f$ represents the flat-field pattern (\citealt{Kameda_2021}). The terms $p_\mathrm{bias}$ and $p_\mathrm{dark}$ denote the bias and dark current intensities, respectively, and $p_\mathrm{hot}$ is the hot pixel pattern. These terms depend on $T_\mathrm{CCD}$, $T_\mathrm{ELE}$, and $T_\mathrm{AE}$, which are the temperatures of the CCD detector, sensor head electronics, and analogue electronics control unit, respectively; see \cite{Tatsumi_2019} for the functional forms of $p_\mathrm{bias}$ and $p_\mathrm{dark}$. Although hot pixel correction has not been included in the nominal data reduction procedure of ONC-T, the significant increase in hot pixels over time, with some reaching intensities comparable to that of the host star, motivated their correction. The hot-pixel patterns were derived by calculating the pixel-wise minimum of two deep-space images acquired at different spacecraft pointings. The reduced images with and without hot pixel correction are compared in Fig. \ref{Fig_oncim}. After correction, the amplitudes of the hot pixels were greatly reduced, remaining well below the level of the stellar signal. In addition, none of the stars in the images were saturated, ensuring that blooming from nearby bright sources did not compromise the photometric analysis.

\begin{figure*}[th]
    \centering
    \includegraphics[width=0.8\hsize]{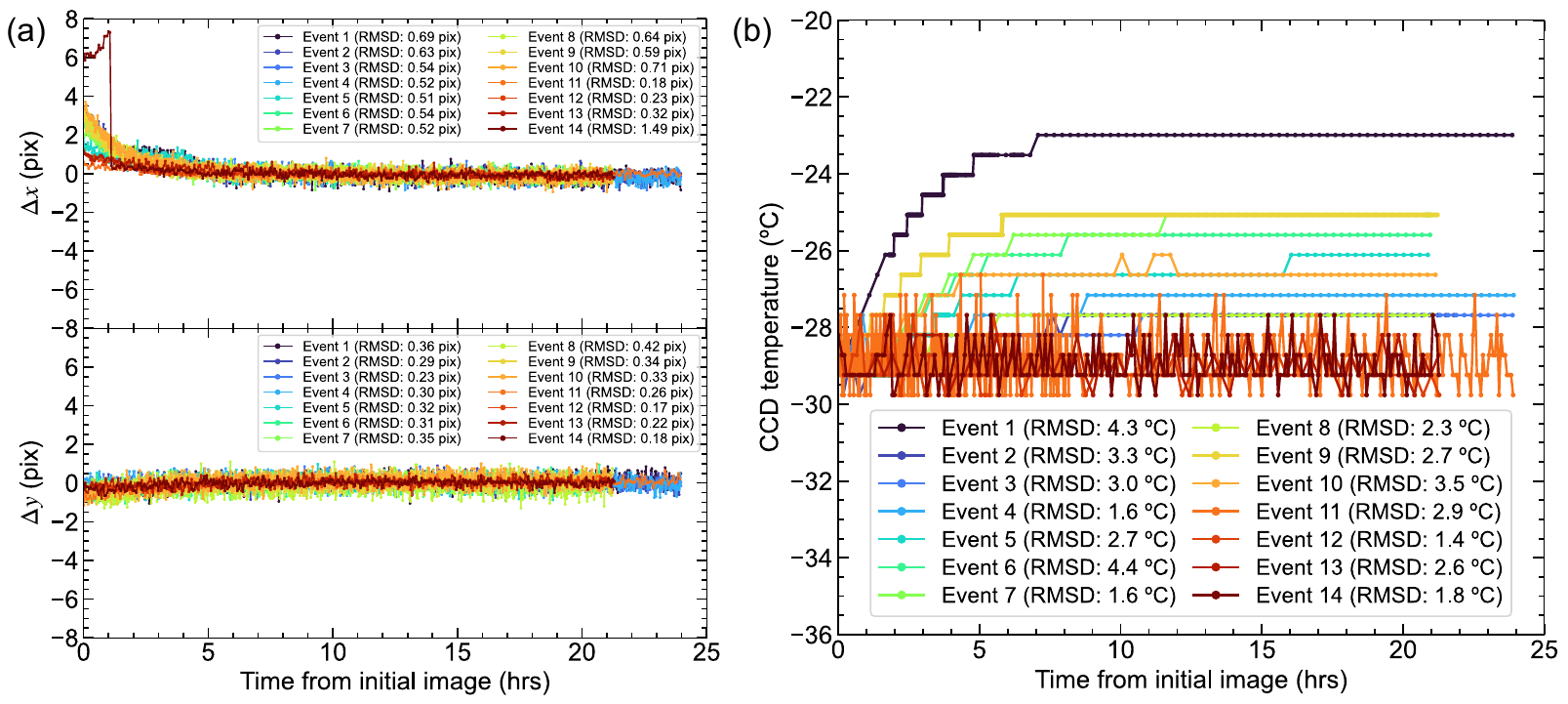}
    \caption{
    Pointing and temperature stability of ONC-T during each transit observation. \textbf{(a)} The top and bottom panels show the horizontal ($x$) and vertical ($y$) centroid positions of the host star, respectively, expressed as deviations from their median values. \textbf{(b)} CCD temperature measured during each observation. The temperature readings are quantized in steps of 0.52°C. The root mean square deviation (RMSD) from the median is indicated in the legend of each panel.
    }
    \label{Fig_oncstability}
\end{figure*}

The light curve ($I^t$) was obtained using aperture photometry:
\begin{equation}
    \label{Eq2}
    I^{t} =
    \sum_{(x,\,y)\,\in\, S_{\mathrm{star}}} p^{t}(x,y)
    \;-\;
    \sum_{(x,\,y)\,\in\, S_{\mathrm{bg}}} p^{t}(x,y).
\end{equation}
Here, $S_\mathrm{star}$ denotes the set of pixel coordinates within the circular aperture used to measure the stellar signal, and $S_\mathrm{bg}$ refers to those within the surrounding annulus for background estimation; both are centered on the stellar centroid. The aperture and annulus sizes were optimized to minimize the RMS of the residuals with respect to the model described in Section \ref{DataRedModel:Model}. Integration was performed over the precise intersection of these regions with each pixel using the \texttt{photutils} library. To mitigate the effects of cosmic rays, a 3$\sigma$ clipping algorithm was applied during background estimation. In addition, data points deviating by more than three times the interquartile range of $I^t$ within a one-hour sliding window were removed.

Since the stellar position drifted over the course of the observations, the centroid was recalculated in each image using a 2D Gaussian fit. In most cases, the star drifted a few pixels in the –$x$ direction during the first five hours and then stabilized to within $\sim$1 pixel (Fig. \ref{Fig_oncstability}a). In Event 14, a sudden jump occurred about one hour into the observation. These positional shifts are attributed to spacecraft pointing instability, as the attitude was adjusted from nominal to the target orientation only 1–2 hours before observations.

A similar drift was observed in the CCD temperature (Fig. \ref{Fig_oncstability}b). Although the heater was set to maintain the CCD between –30 and –28°C, the temperature typically rose during the first $\sim$8 hours of each observation, reaching as high as –23°C in the most extreme case before asymptotically stabilizing.

For the $I^t$ of TESS data, we used the Pre-search Data Conditioning Simple Aperture Photometry (PDCSAP) flux, provided at a 120-second cadence and processed by the TESS Science Processing Operations Center at NASA Ames Research Center (\citealt{Jenkins_2016}). For each transit event, data were extracted within a 26-hour window centered on the predicted mid-transit time, although coverage within this window was not always continuous. Data points flagged with quality values of 1, 2, 4, 8, 32, or 128 were excluded (\citealt{Twicken_2020}).

\subsection{Transit modeling and noise decorrelation}
\label{DataRedModel:Model}

For each transit event, the light curve $I^t$ was normalized by its median to obtain $\widehat{I^{t}}$, and then fit with a model combining the transit signal ($\mathcal{T}(t)$) and a systematic noise component ($\mathcal{N}(t)$):
\begin{equation}
    \label{Eq3}
    \widehat{I^{t}} = \mathcal{T}(t) + \mathcal{N}(t).
\end{equation}
For $\mathcal{T}(t)$, we used the \cite{Mandel_Agol_2002} transit model as implemented in the \texttt{PyTransit} library (\citealt{Parviainen_2015}), assuming a circular orbit and a quadratic limb-darkening law. The out-of-transit baseline of $\mathcal{T}(t)$ equals one. Following the approach used in previous TTV studies (e.g., \citealt{Kokori_2022}; \citealt{Ivshina_2022}), we fitted the mid-transit time ($T_C$) and the planet-to-star radius ratio ($R_p/R_\ast$), while fixing the orbital period ($P$), semi-major-axis–to–star-radius ratio ($a/R_\ast$), orbital inclination ($i_p$), and quadratic limb-darkening coefficients ($u_1$, $u_2$) to literature values: $P=$2.7240308 days (\citealt{Ivshina_2022}), $a/R_\ast$=4.6, $i_p$=84.03$^\circ$, $u_1$=0.495, $u_2$=0.055 (\citealt{Lendl_2020}) for WASP-189 b and $P=$2.14877381 days, $a/R_\ast$=4.1676, $i_p$=88.45$^\circ$, $u_1$=0.392, $u_2$=0.092 (\citealt{Hooton_2022}) for MASCARA-1 b. Since the fitting was performed individually for each transit event, the assumed orbital period influences the fit only through its effect on the transit duration, not on the mid-transit time. A precise value of the orbital period is later derived from the compiled set of $T_C$ (Section \ref{Results:TTV}).

For the ONC-T data, pixel-level decorrelation (PLD) model was used as $\mathcal{N}(t)$ to account for spacecraft pointing instabilities:
\begin{align}
    \label{Eq4}
    \mathcal{N}(t) &=
    \sum\limits_{(x_i,y_i)\in M}
      c(x_i,y_i)\,
      \frac{p^{t}(x_i,y_i)}
           {\sum\limits_{(x_j,y_j)\in M} p^{t}(x_j,y_j)} \notag \\
    &\qquad + \alpha + \beta\,(t - t_{0})
      + \eta \left[ 1 - \exp\!\left(-\tfrac{t - t_{0}}{\tau}\right) \right].
\end{align}
The first term represents the first-order PLD model \citep{Deming_2015}, while the remaining terms account for time-correlated noise. The set of pixel coordinates used as the PLD basis vector is denoted by $M$, and each of their coefficients are denoted by $c$. The time-correlated component is modeled as a combination of a linear trend and an exponentially decaying function that asymptotically approaches zero. Here, $t_0$ denotes the start time of observation. The decaying function is intended to model noise correlated with CCD temperature, which follows a similar temporal pattern (Fig. \ref{Fig_oncstability}b). The fitting parameters are $c(x_i,y_i)$ for $i=1,2,\ldots,|M|$, as well as $\alpha$, $\beta$, $\eta$, and $\tau$. For the PLD basis vectors, we selected pixels within a 3 × 3 window centered on the mean stellar centroid, supplemented by any additional pixels the centroid crossed in individual frames. This resulted in a total of $|M| =$ 9 to 15 pixels, depending on the degree of pointing stability. This selection strategy effectively enabled the PLD model to capture pointing drift during the initial phase of the observation, while avoiding an excessive increase in the number of model parameters.

For the TESS data, a third order polynomial was used as $\mathcal{N}(t)$:
\begin{equation}
    \label{Eq5}
    \mathcal{N}(t) =
    \alpha \;+\; \beta (t - t_{0})
    \;+\; \gamma (t - t_{0})^{2}
    \;+\; \delta (t - t_{0})^{3}.
\end{equation}
Here, the coefficients $\alpha$, $\beta$, $\gamma$, and $\delta$ are the fitting parameters.

The parameter spaces of both $\mathcal{T}(t)$ and $\mathcal{N}(t)$ were jointly explored using the Markov Chain Monte Carlo (MCMC) sampling implemented in the \texttt{emcee} library (\citealt{Foreman-Mackey_2013}). We adopted uniform priors for all parameters and ran the MCMC with 1,024 walkers for 10,000 steps, discarding the first 5,000 as burn-in. The best-fit parameter values and their 1$\sigma$ credible intervals were obtained as the median and standard deviation of the marginalized posterior distributions, respectively.

\subsection{Blind search of the exoplanet using Transit Least Squares}
\label{DataRedModel:TLS}

While the modeling described in Section \ref{DataRedModel:Model} assumes the presence of a transit signal, we also applied an alternative approach that makes no such assumption to test whether the transit signal can be recovered solely from ONC-T data. In this approach, we set $\mathcal{T}(t)=1$ (no transit model) and fit the data using only $\mathcal{N}(t)$.

A periodic transit signal was searched for in the noise-detrended light curves ($\widehat{I^t}-\mathcal{N}(t)$) of all events using the Transit Least Squares (TLS) algorithm implemented in the \texttt{transitleastsquares} library \citep{Hippke_2019}. TLS has been widely used to find transit signals in Kepler and TESS data (e.g., \citealt{VanGrootel_2021}; \citealt{Lillo-Box_2020}). The method phase-folds the light curve over a range of trial periods and computes the signal detection efficiency (SDE), which quantifies the statistical significance of a transit at each trial period. The false-positive rate of transit detections drops below 1\% when the SDE exceeds 7. 

% -------------------------------------- %
% ------------- Results ---------------- %
% -------------------------------------- %

\section{Results: Successful transit detections by ONC-T}
\label{Results}

\begin{figure*}[p]
    \centering
    \includegraphics[width=0.88\hsize]{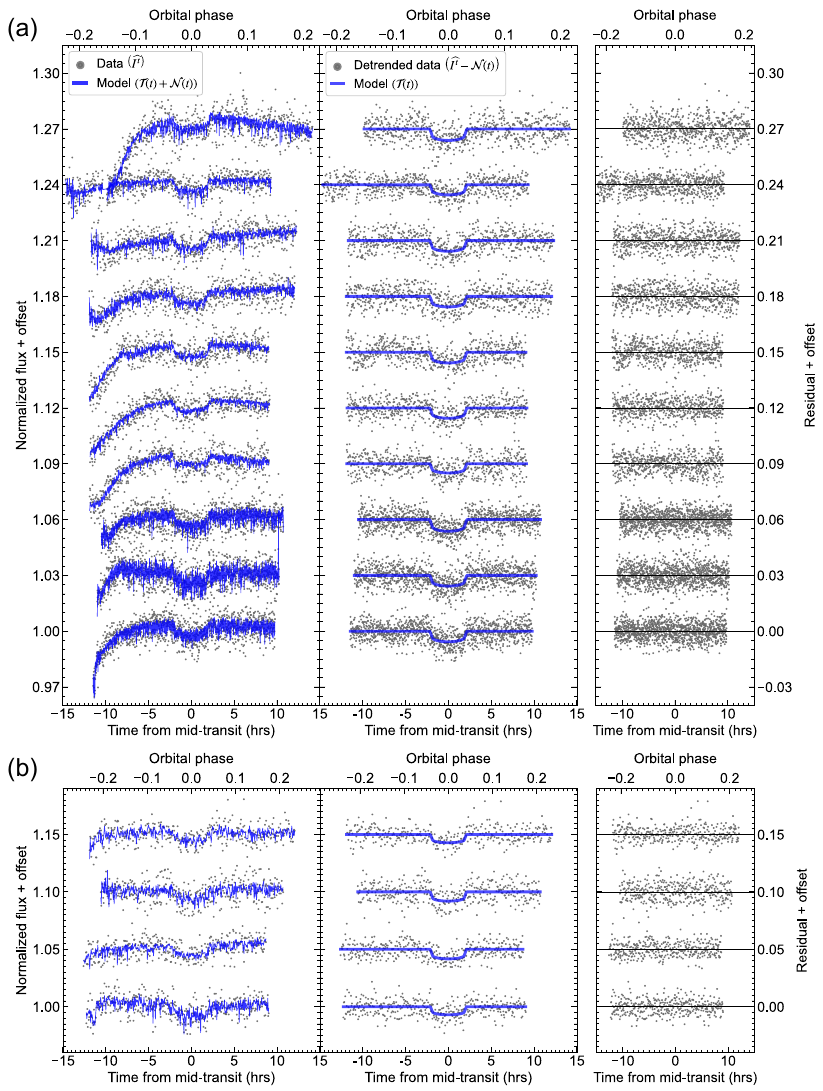}
    \caption{
    Light curves for each transit of \textbf{(a)} WASP-189 b and \textbf{(b)} MASCARA-1 b observed by ONC-T. Left panels show the raw light curves ($\widehat{I^t}$), with blue curves representing the combined transit and noise model fit ($\mathcal{T}(t)+\mathcal{N}(t)$). Center panels display the noise-detrended light curves ($\widehat{I^t}-\mathcal{N}(t)$), with blue curves showing the fitted transit model ($\mathcal{T}(t)$). Right panels show the residuals: $\widehat{I^t}-[\mathcal{T}(t)+\mathcal{N}(t)]$. Light curves from transit events 1–10 for WASP-189 b and 11–14 for MASCARA-1 b are displayed from top to bottom in each panel. Each light curve is vertically offset for clarity by 0.03 in panel (a) and 0.05 in panel (b).
    }
    \label{Fig_lc}
\end{figure*}

\begin{figure*}[th]
    \centering
    \includegraphics[width=0.85\hsize]{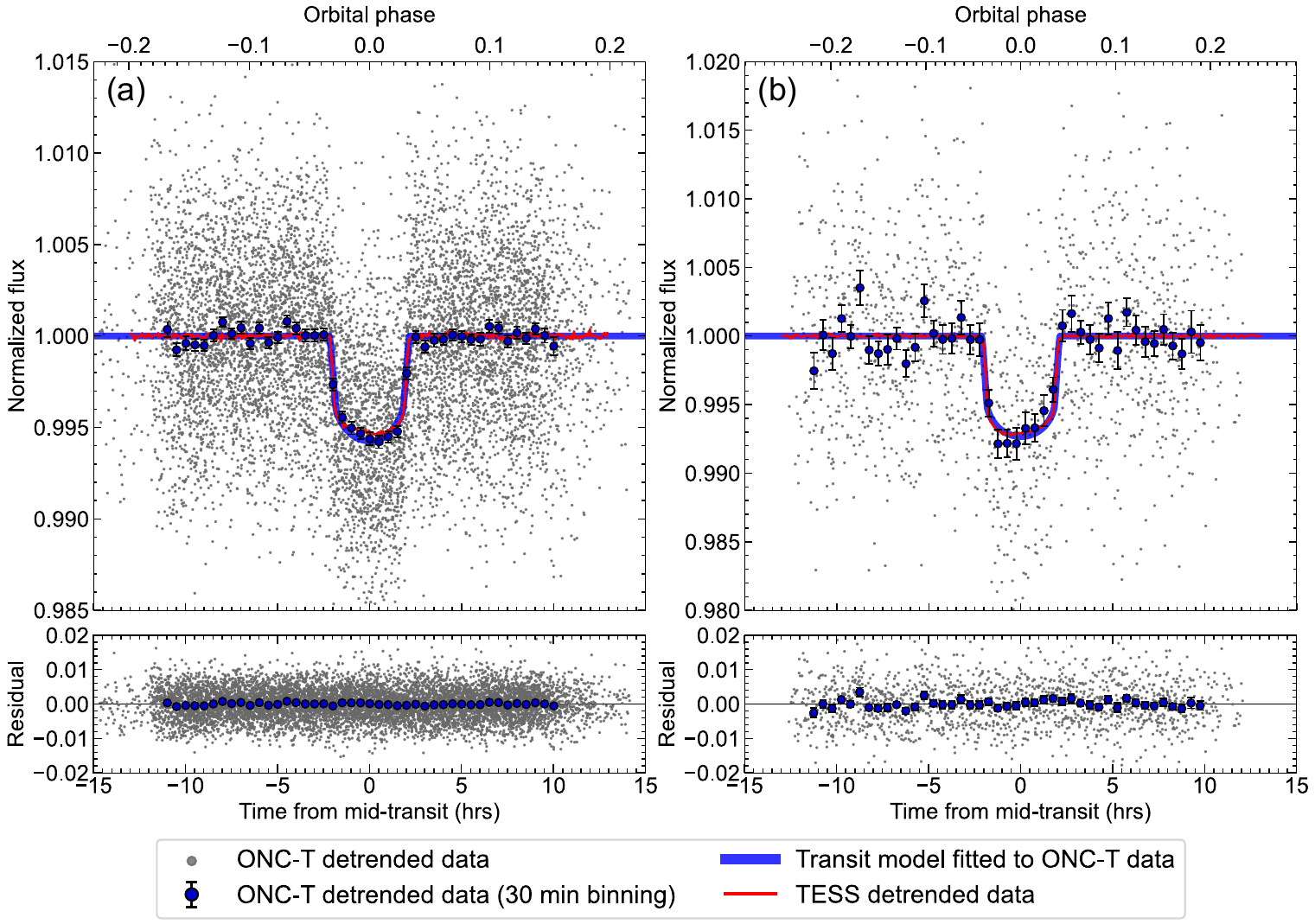}
    \caption{
    Phase-folded light curves of \textbf{(a)} WASP-189 b and \textbf{(b)} MASCARA-1 b from ONC-T and TESS data. The top panels show the phase-folded detrended light curves ($\widehat{I^t}-\mathcal{N}(t)$), while the bottom panels display the residuals of the ONC-T data relative to the fitted transit model ($\mathcal{T}(t)$; solid blue curve). Gray points show the flux measured in individual ONC-T images, while blue circles represent 30-minute binned averages, with error bars indicating the standard error. Error bars are not visible when smaller than the marker size. The red curve represents the TESS light curve, baseline-corrected and binned over 4-minute intervals.
    }
    \label{Fig_foldedlc}
\end{figure*}

\begin{table*}[b]
\centering
\caption[]
{
\label{Table_ephemeris}
Updated reference epoch ($T_{0}$) and orbital period ($P$) compared with values reported in the latest literature.
}
    
\setlength\tabcolsep{5.5pt} % column spacing
\renewcommand{\arraystretch}{1.2} % row spacing
    
\begin{tabular}{l l l}
\hline\hline
 & $T_{0}$ (BJD$_{\mathrm{TDB}}$) & $P$ (days) \\
\hline
\multicolumn{3}{c}{WASP-189 b}\\ \hline 
Updated (This study)             & $2456706.456164 \pm 0.000313$ & $2.72403141 \pm 0.00000035$ \\
Literature (\citealt{Ivshina_2022}) & $2456706.456600 \pm 0.002300$ & $2.72403080 \pm 0.00000280$ \\
\hline
\multicolumn{3}{c}{MASCARA-1 b}\\ \hline
Updated (This study)            & $2458833.488125 \pm 0.000079$ & $2.14876998 \pm 0.00000013$ \\
Literature (\citealt{Hooton_2022}) & $2458833.488151^{+0.000091}_{-0.000092}$ & $2.14877381^{+0.00000088}_{-0.00000087}$ \\
\hline
\end{tabular}
    
\end{table*}

\subsection{Statistics of transit light curves observed by ONC-T}
\label{Results:Statistics}

The ONC-T data reveal statistically significant transit signals from both WASP-189 b and MASCARA-1 b. The raw light curve ($\widehat{I^t}$; Fig. \ref{Fig_lc} left) exhibits a rising baseline during the first $\sim$8 hours, likely due to drifts in CCD temperature and spacecraft pointing (Fig. \ref{Fig_oncstability}). Despite this, the transit dip extending approximately $\pm$2 hours from mid-transit is clearly visible. After subtracting the systematic noise ($\widehat{I^t}-\mathcal{N}(t)$; Fig. \ref{Fig_lc} center), the baseline flattens, making the transit more distinct. The photometric precision per ONC-T image is calculated as 0.46 $\pm$ 0.05\% for WASP-189 b and 0.65 $\pm$ 0.06\% for MASCARA-1 b, based on the RMS of the residuals: $\widehat{I^t}-[\mathcal{T}(t)+\mathcal{N}(t)]$ (Fig. \ref{Fig_lc} right). We assess the significance of the transit signal by calculating its SNR using the following equation:
\begin{equation}
    \label{Eq6}
    \mathrm{Transit\;SNR} \;=\; 
    \frac{T_{\mathrm{dep}}}{\sqrt{\sigma_{\mathrm{in}}^{2} + \sigma_{\mathrm{out}}^{2}}}.
\end{equation}
Here, $T_\mathrm{dep}$ is the transit depth, $\sigma_{\mathrm{in}}$ is the standard error of the in-transit flux, and $\sigma_{\mathrm{out}}$ is the standard error of the out-of-transit flux. The transit SNR evaluated for a single transit event is 12.5 $\pm$ 2.7 for WASP-189 b and 7.9 $\pm$ 0.7 for MASCARA-1 b.

After phase folding the 10 transit observations of WASP-189 b and four of MASCARA-1 b, their transit SNR increase to 39.8 and 15.6, respectively (Fig. \ref{Fig_foldedlc}). The shapes of the phase-folded light curves closely resemble those observed by TESS. The TESS data exhibit asymmetric transit curves, attributed to gravity darkening induced by the host star’s rapid rotation (\citealt{Lendl_2020}; \citealt{Hooton_2022}). A similar asymmetry is visible in the ONC-T data, although its magnitude is comparable to the uncertainty.

The posterior distributions of the derived $T_C$ and $R_p/R_*$ follow unimodal Gaussian shapes and show negligible mutual correlation (correlation coefficient $<0.04$; Appendix \ref{Appendix:Posterior}). The 1$\sigma$ precision in $T_C$ derived from a single transit is 5.4 min for WASP-189 b and 6.9 min for MASCARA-1 b. The 1$\sigma$ precision in $R_p/R_*$ is 0.0036 (5.1\% relative) for WASP-189 b and 0.0062 (7.7\% relative) for MASCARA-1 b. A higher precision was achieved in events 8–10 of WASP-189 b, owing to a roughly twofold reduction in the imaging cadence by minimizing overhead duration (Table \ref{Table_obscond}).

\begin{figure*}[th]
    \centering
    \includegraphics[width=0.97\hsize]{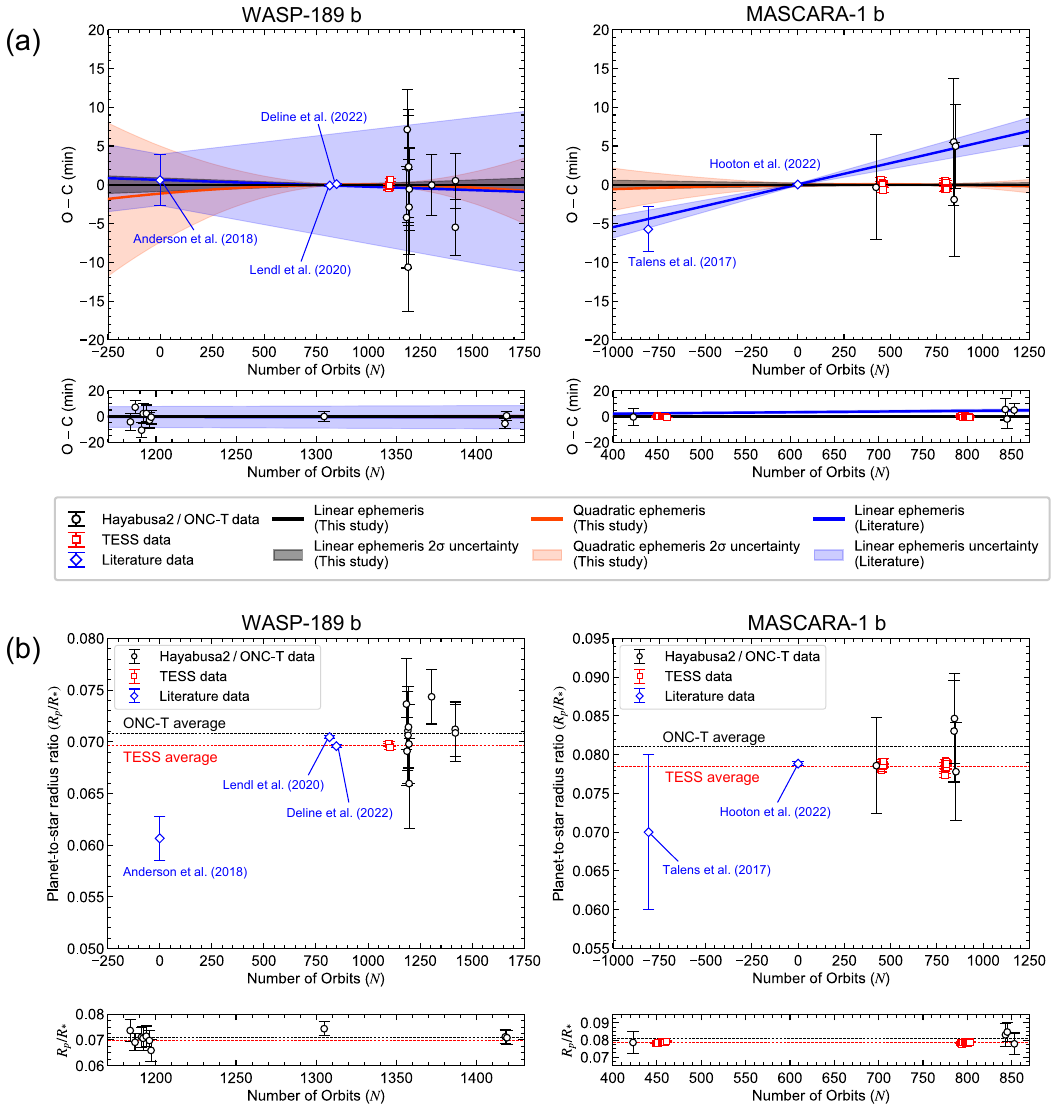}
    \caption{
    \textbf{(a)} Transit timing and \textbf{(b)} planet-to-star radius ratio ($R_p/R_*$) variations of \textbf{(left)} WASP-189 b and \textbf{(right)} MASCARA-1 b. The O – C values on the vertical axis in (a) represent the difference between the observed transit times and those computed by the linear ephemeris model derived in this study. The fitted linear and quadratic models are shown as black and red lines, respectively, with their 2$\sigma$ uncertainty regions indicated by hatching. For comparison, the linear model reported in the most recent literature is shown in blue, with its reported uncertainty shaded. For each figure in (a) and (b), the bottom panels provide a magnified view of the ONC-T observation period. Data points from the literature are labeled with their respective references.
    }
    \label{Fig_TTV_TDepV}
\end{figure*}

\subsection{Transit timing analyses}
\label{Results:TTV}

We fit the following linear ephemeris model, assuming a constant orbital period, to all $T_C$ values derived from ONC-T, TESS, and previous observations:
\begin{equation}
    \label{Eq7}
    T_{C}(N) = T_{0} + NP.
\end{equation}
Here, $N$ is the orbit number and $T_0$ is the reference epoch at $N=0$. Using this model, we generate an O – C diagram (Fig. \ref{Fig_TTV_TDepV}a) that shows the deviation of the observed $T_C$ values from those computed by the fitted linear ephemeris. We also test to fit the data with a quadratic ephemeris model that accounts for a constant change in orbital period due to orbital decay (\citealt{Patra_2017}):
\begin{equation}
    \label{Eq8}
    T_{C}(N) = T_{0} + NP + \frac{1}{2}\,\frac{dP}{dN}\,N^{2}.
\end{equation}
Here, $dP/dN$ represents the orbital decay rate. The model parameters and their uncertainties are derived using MCMC with uniform priors, following the same procedure described in Section \ref{DataRedModel:Model}.

The $T_C$ derived from ONC-T data show no systematic deviation from the fitted linear ephemeris model (Fig. \ref{Fig_TTV_TDepV}a). The average O – C values are –1.2 min for WASP-189 b and +2.1 min for MASCARA-1 b, indicating that measurement accuracy of $T_C$ is better than $\sim$2 min. 

The new transit data from TESS and ONC-T enabled us to refine the ephemeris models (Table \ref{Table_ephemeris}). For WASP-189 b, the uncertainties in $T_0$ and $P$ were reduced by a factor of $\sim$6, while remaining consistent with the values reported by \cite{Ivshina_2022} within their uncertainties. All reported transit timings show no significant deviation from the linear ephemeris model. The orbital decay rate $dP/dN$ derived from the quadratic model is $-1.8 \pm 4.7 \times 10^{-9}$ days, indicating no significant change in the orbital period over the past 10.6 years, from Feb 2014 to Sep 2024. 

\begin{figure*}[t!]
    \centering
    \includegraphics[width=0.9\hsize]{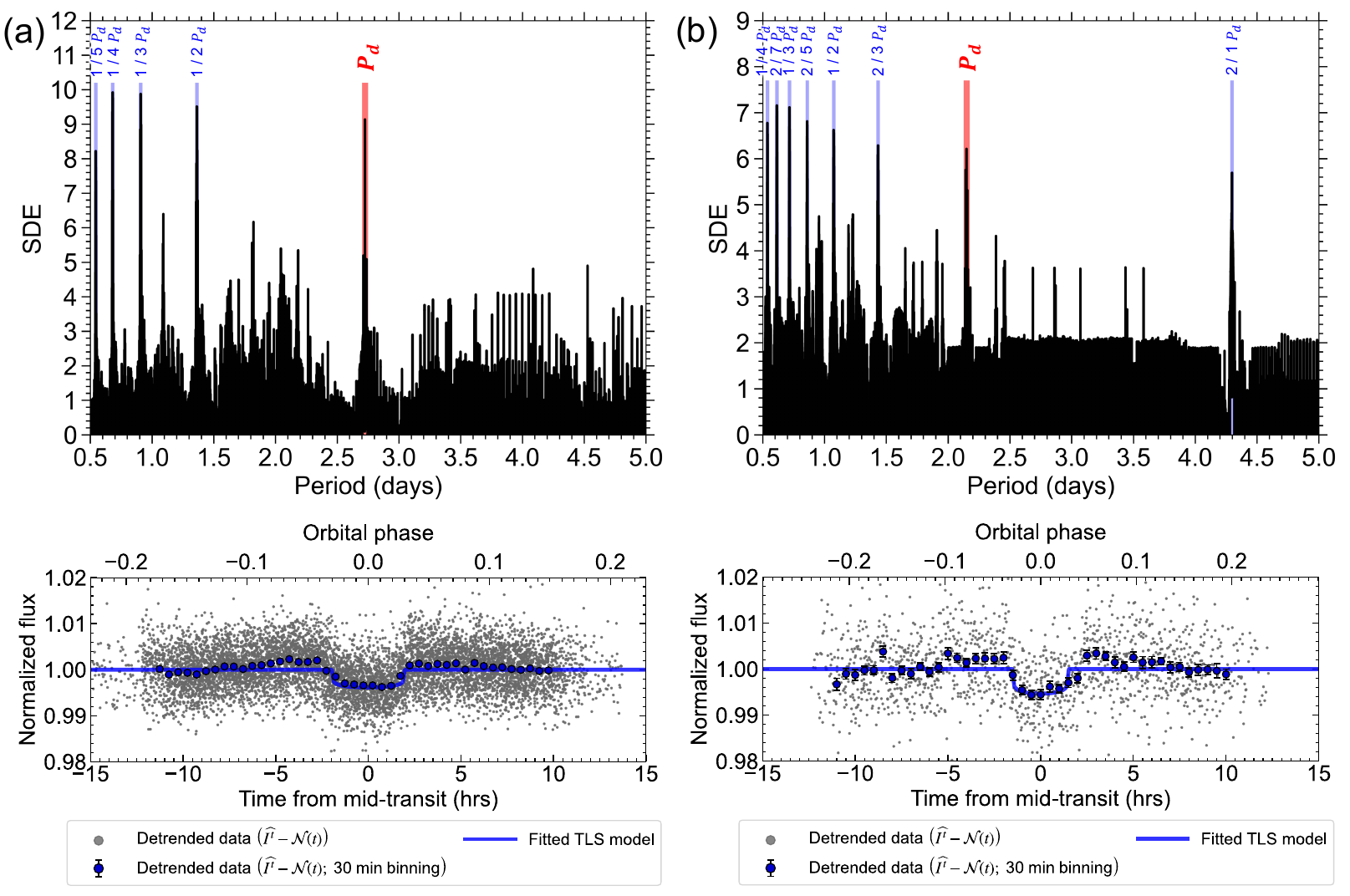}
    \caption{
    TLS analysis results for \textbf{(a)} WASP-189 b and \textbf{(b)} MASCARA-1 b. The top panels show the TLS periodograms, with the red vertical line marking the detected period ($P_d$), consistent with the true orbital period ($P$). Blue lines indicate additional peaks with SDE $>$ 6 for WASP-189 b and SDE $>$ 4 for MASCARA-1 b; all correspond to rational multiples or fractions of $P_d$. The bottom panels display the detrended light curves ($\widehat{I^t}-\mathcal{N}(t)$) phase-folded at $P_d$. Gray points show the flux measured in individual ONC-T images, while blue circles represent 30-minute binned averages, with error bars indicating the standard error. Error bars are not visible when smaller than the marker size. The solid blue curve is the best-fit TLS model.
    }
    \label{Fig_TLS}
\end{figure*}

In contrast, for MASCARA-1 b, we find a significant discrepancy of 4$\sigma$ between our refined ephemeris model and those reported in the literature (\citealt{Hooton_2022}; Table \ref{Table_ephemeris}). \cite{Hooton_2022} derived the orbital period based on observations of three transits (two from CHEOPS and one from Spitzer/IRAC at 4.5 µm) and four occultations (two each from CHEOPS and Spitzer). This ephemeris aligns better with earlier transit timings reported by \cite{Talens_2017}, but deviates significantly from recent TESS observations by up to 4.4 $\pm$ 0.4 min. Conversely, the ephemeris model derived in this study is consistent with recent TESS data, yet shows a discrepancy of 6 $\pm$ 3 min with the \cite{Talens_2017} timings. Including an orbital decay term does not resolve this discrepancy; the derived $dP/dN$ of $-4.1 \pm 9.6 \times 10^{-10}$ days indicates no constant change in the orbital period. The cause of this discrepancy may require further investigation through continued monitoring of transit timings and by considering models that account for other sources of TTVs. The ONC-T data acquired to date are consistent with both the refined and literature ephemerides within the errors. Nevertheless, ONC-T is expected to distinguish between the two models with 2$\sigma$ significance by 2030, during which Hayabusa2 is planned to still be in cruise. We recommend using the updated ephemeris derived in this study for planning future transit observations, as it provides a better match to recent data.

\subsection{Transit depth analyses}
\label{Results:Tdep}

The $R_p/R_*$ values derived from ONC-T data agree with those from TESS data within 0.0011 (1.6\% relative) for WASP-189 b and 0.0025 (3.1\% relative) for MASCARA-1 b, supporting that the ONC-T measurements are accurate to at least these levels (Fig. \ref{Fig_TTV_TDepV}b).

The $R_p/R_\ast$ values derived from recent ONC-T and TESS observations are generally consistent with past measurements, showing no significant changes over time. The relatively low $R_p/R_\ast$ of WASP-189 b reported by \citet{Anderson_2018} has been attributed to potential inaccuracies due to the lack of full ingress-to-egress coverage in their follow-up observations and the scarcity of nearby reference stars \citep{Lendl_2020}. Our results further support this interpretation. For MASCARA-1 b, \citet{Talens_2017} reported large uncertainties in $R_p/R_\ast$, caused by discrepancies between measurements from the MASCARA and NITES telescopes: $R_p/R_\ast= 0.063_{-0.001}^{+0.002}$ from MASCARA data and $0.078_{-0.001}^{+0.002}$ from NITES data. Our results align with the NITES-based estimate.

\subsection{Transit Least Squares analyses}\label{Results:TLS}

The TLS analysis demonstrates successful true-positive detections of the exoplanets, indicating that ONC-T data alone provide sufficient information for their detection. The TLS periodograms (Fig. \ref{Fig_TLS}) show a significant peak near the actual orbital periods of the exoplanets. The period of the detected peak ($P_d$) is 2.72392 $\pm$ 0.00048 days for WASP-189 b and 2.14877 $\pm$ 0.00031 days for MASCARA-1, where the uncertainties correspond to the half-width at half-maximum. These periods are consistent with the actual orbital periods within their errors (Table \ref{Table_ephemeris}). The SDE at $P_d$ is 9.1 for WASP-189 b and 6.2 for MASCARA-1 b. The SDE for WASP-189 b exceeds the commonly adopted threshold of 7, corresponding to a false-positive rate below 1\% (\citealt{Hippke_2019}). Although the SDE for MASCARA-1 b falls below this threshold, likely due to its fainter host star and fewer observed transits, it can still be considered a strong planetary candidate, as some studies adopt an empirical threshold of SDE $>$ 6 (e.g., \citealt{Dressing_2015}).

All other peaks with SDE values comparable to that at $P_d$ correspond to rational multiples or fractions of $P_d$. This pattern arises from aliasing effects due to the non-continuous temporal coverage of ONC-T observations. Therefore, even peaks with higher SDEs than at $P_d$ do not necessarily indicate false positives. Although the risk of false-positive detections cannot be entirely ruled out, since the observation windows were intentionally aligned with the expected transit times, the flat out-of-transit flux profile suggests that false positives are unlikely to be a significant concern.

In addition, the TLS analysis confirms that $\mathcal{N}(t)$ does not overfit the transit signal. When $\mathcal{T}(t)$ is omitted and the data are fitted using only $\mathcal{N}(t)$, the resulting light curves exhibit a slightly wing-shaped baseline (Fig.~\ref{Fig_TLS}, bottom), in contrast to the cleaner baselines obtained when $\mathcal{T}(t)$ is included (Fig.~\ref{Fig_foldedlc}). Nevertheless, the transit shape is well preserved, supporting the reliability of our denoising approach.

% ------------------------------------------ %
% ------------- Discussions ---------------- %
% ------------------------------------------ %

\section{Discussion: Implications for future small satellite observations}
\label{Discussion:Implications}

Building on the benchmark established by ONC-T, we derive a formula to estimate the transit detectability by future missions. Assuming photon shot noise is the dominant noise source, the transit SNR (Equation \ref{Eq6}) can be estimated for a wide range of targets and instrument parameters using the following formula (\citealt{Kokori_2022,Kokori_2023}):
\begin{align}
    \label{Eq9}
    &\mathrm{Predicted\;SNR} \;=\; \notag\\
    &Q \cdot D \cdot T_{\mathrm{dep}}
    \cdot \sqrt{10^{-m/2.5}}
    \;\left(
      \frac{1}{T_{\mathrm{dur}}}
      + \frac{1}{T - T_{\mathrm{dur}}}
    \right)^{-\tfrac{1}{2}}.
\end{align}

Here, $D$ is the aperture diameter in millimeters, $m$ is the apparent magnitude of the host star at the observed wavelength, $T_\mathrm{dep}$ is the transit depth in percent, $T_\mathrm{dur}$ is the transit duration in hours, and $T$ is the total observation time in hours. The scaling factor $Q$ is an empirical constant derived from ONC-T observations. We find $Q=20.88$, under which the predicted transit SNR for MASCARA-1 b ($m=8.3$, $T_\mathrm{dep}=0.62$\%, and $T_\mathrm{dur}=4.2$ hours) matches the observed SNR of 7.9 under ONC-T’s observation conditions ($D=15$ mm; $T=21$ hours). Based on the empirical relation between the transit SNR and SDE found in this study, a transit SNR above $\sim$10 is likely required to achieve SDE $>$ 7 and confidently detect a new exoplanet.

We confirm that Equation \ref{Eq9} predicts transit SNRs with reasonable accuracy. For instance, the formula predicts a transit SNR of 15.0 for WASP-189 b observed by ONC-T, in good agreement with the observed SNR (Section \ref{Results:Statistics}). For the TESS observations, the predicted SNRs are 105 for WASP-189 b and 55 for MASCARA-1 b, both consistent with the observed values of 130 and 70, respectively, within 30\%. The slight underestimation likely results from ONC-T’s lower pointing and thermal stability compared to TESS (Table \ref{Table_review}), along with longer overhead durations. For 55 Cnc e observed by ASTERIA, the predicted SNR of 3.8 agrees within a factor of two with the observed SNR of 2.2 (\citealt{Knapp_2020}). The observed SNR is lower in this case likely due to contamination from the nearby star 53 Cnc or increased pixel-to-pixel gain variations inherent to ASTERIA’s CMOS design.

\begin{figure}[th]
    \centering
    \includegraphics[width=1.0\hsize]{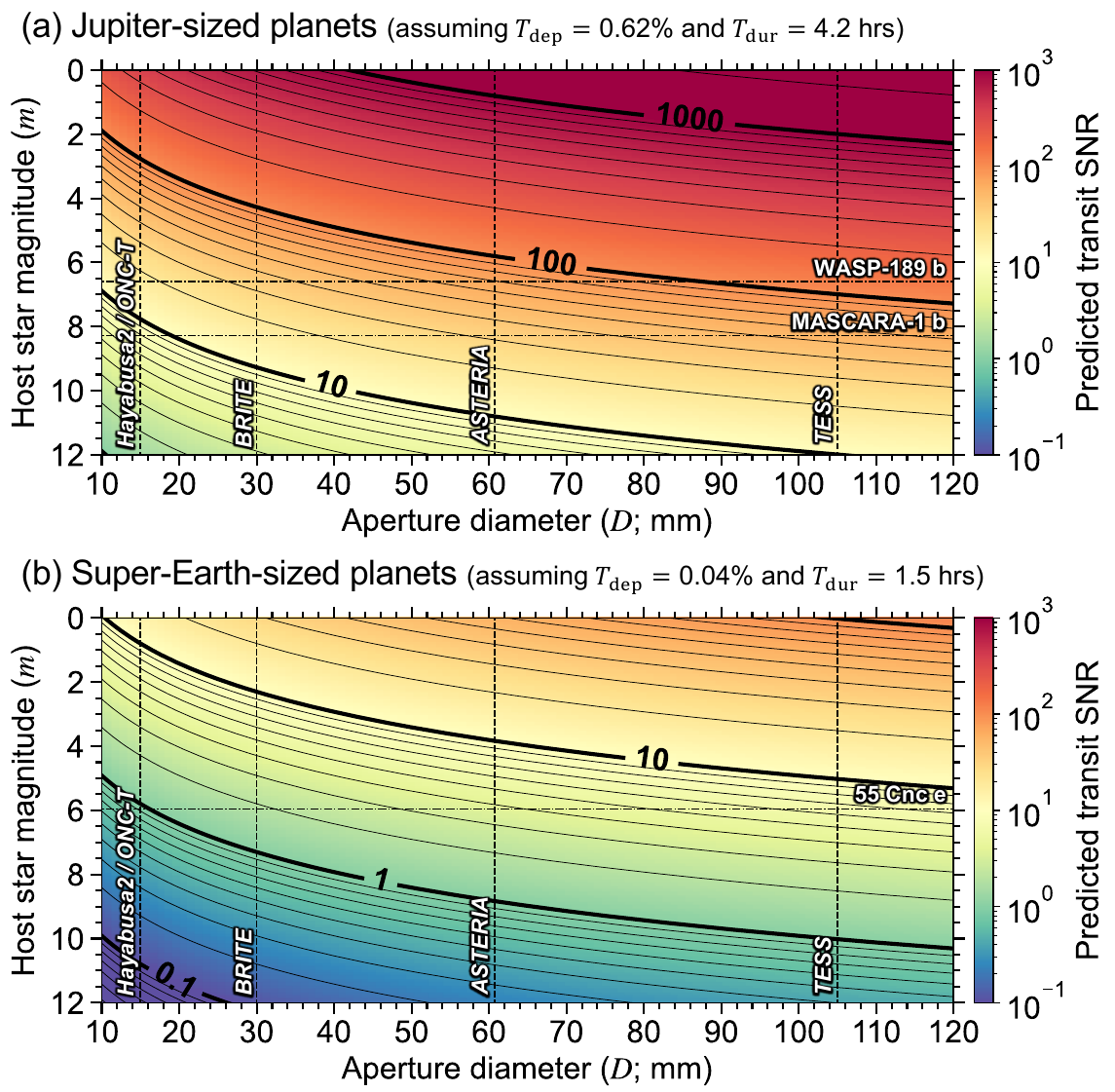}
    \caption{
    Predicted transit signal-to-noise ratio (SNR) for \textbf{(a)} Jupiter-sized and \textbf{(b)} super-Earth-sized planets, shown as a function of telescope aperture diameter ($x$-axis) and host star magnitude ($y$-axis). Vertical dashed lines indicate the apertures of past space-based instruments. Horizontal lines mark the host star magnitudes of WASP-189 b and MASCARA-1 b in panel (a), and 55 Cnc e in panel (b). For the Jupiter-sized planet case, the transit depth ($T_\mathrm{dep}$) and duration ($T_\mathrm{dur}$) of MASCARA-1 b were assumed in the SNR prediction. For the super-Earth-sized planet case, the corresponding values of 55 Cnc e were used.
    }
    \label{Fig_transitSNR}
\end{figure}

We use Equation \ref{Eq9} to estimate transit SNRs across a range of $D$ and $m$ conditions (Fig. \ref{Fig_transitSNR}). For instruments with aperture sizes as small as that of ONC-T, the predicted SNR of Jupiter-sized planets exceeds 10 when the host star is brighter than $m\!\sim\!7.5$. More than 25,000 stars across the sky surpass this brightness threshold (\citealt{Hog_2000}). Assuming each of these stars hosts a single Jupiter-sized planet with an orbital period uniformly distributed in $\log_{10}(\mathrm{Period})$ from 1 day to 100 years, we estimate that approximately 40, 15, and 3 transiting Jupiters would be detectable for orbital periods longer than 100 days, 1 year, and 10 years, respectively
\footnote{We assume the geometric transit probability
$(R_{\ast}+R_{p})/a$ to be
\[
\frac{R_{\odot}+R_{J}}{1\,\mathrm{AU}}
\left( \frac{P}{365\,\mathrm{days}} \right)^{-2/3},
\]where \(R_{\odot}\) and \(R_{J}\) are the solar and Jovian radii, respectively, expressed in astronomical units (AU). This expression assumes a circular orbit, a host star with solar mass and radius, and a planet with Jupiter’s radius.}.
Thus, our results suggest that long-term photometric monitoring of bright stars using small satellites equipped with ONC-T-class optics could play a valuable role in surveying the population of long-period ($>$100 days) transiting planets, which remain underrepresented in current surveys (Fig. \ref{Fig_exoplanets}b). In contrast, detecting super-Earth-sized planets with similar instrumentation requires much brighter host stars. For ONC-T-class optics, achieving an SNR greater than 10 is only possible for stars brighter than about $m\sim$1, of which there are only around 10. This suggests that larger apertures are necessary to gather statistically meaningful data on super-Earth populations. For instance, increasing the aperture to 60 mm would raise the brightness threshold to $m\sim$4 and expand the number of stars to over 500.

% ------------------------------------------ %
% ------------- Conclusions ---------------- %
% ------------------------------------------ %
\section{Conclusions}
\label{Conclusions}
We reported the successful transit detections of two known hot Jupiters, WASP-189 b and MASCARA-1 b, using the telescopic Optical Navigation Camera (ONC-T) onboard the Hayabusa2 spacecraft. With its effective aperture diameter of 15 mm, ONC-T sets a new record for the smallest aperture optics to detect an exoplanet transit from space. Each transit event of WASP-189 b and MASCARA-1 b were detected with signal-to-noise ratios (SNR) of 13 and 8, respectively. Combining all observed transits (ten for WASP-189 b, four for MASCARA-1 b) increased the SNR to 40 and 16. A blind search for transit signals using the Transit Least Squares (TLS) algorithm successfully recovered periodic signals with signal detection efficiencies (SDEs) exceeding 9 for WASP-189 b and 6 for MASCARA-1 b.

Transit mid-times ($T_C$) for WASP-189 b and MASCARA-1 b were measured with precisions of 5 and 7 minutes, respectively, per single transit. Their planet-to-star radius ratios ($R_p/R_*$) were measured with precisions of 0.004 (5\% relative) and 0.006 (8\% relative), respectively, per transit. The $T_C$ and $R_p/R_*$ derived from ONC-T data agree with those from TESS within accuracies better than 2 minutes and 0.002 (3\% relative), respectively. Combining all $T_C$ data from ONC-T, TESS, and literatures, we reduced the uncertainties in the ephemerides of both planets. Notably, the updated orbital period of MASCARA-1 b (2.14876998 $\pm$ 0.00000013 days) differs from previous literature by 4$\sigma$, suggesting possible transit timing variations and highlighting the need for continued monitoring.

The successful transit detections and characterizations by ONC-T demonstrate that compact optics, such as those deployable on CubeSats and nanosatellites, can play a significant role in advancing exoplanet population studies. Using ONC-T as a benchmark, we developed a formula to predict transit SNR across various target brightnesses and aperture sizes. One promising application of optics as small as ONC-T’s is the search for transiting, long-period ($>$100 days) Jupiter-sized planets—a population still underrepresented in current catalogs—through long-term monitoring of stars brighter than approximately magnitude 7.5.

\section{Data}
All the {\it TESS} data used in this paper can be found in MAST: \dataset[10.17909/t9-nmc8-f686]{http://dx.doi.org/10.17909/t9-nmc8-f686}. Data presented in Fig. \ref{Fig_exoplanets} can be found in the NASA Exoplanet Archive: \dataset[10.26133/NEA1]{[https://doi.org/10.26133/NEA1}.

%% Please use the acknowledgment and contribution environments. This will 
%% be anonomyized when the "anonymous" style option is used. 
\begin{acknowledgements}

We are grateful to the Hayabusa2 team for their years of dedicated operations and for making the exoplanet observations possible. We deeply appreciate Dr. Rie Honda’s commitment to the development, data archiving, and operations of the Optical Navigation Camera onboard Hayabusa2. This work was supported by Grant-in-Aid for JSPS Fellows Grant number 24KJ2228. 

\end{acknowledgements}

\begin{contribution}
%%This section gives authors the space to recognize author contributions. The text inside this environment is NOT counted towards the total word quanta. At a minimum, manuscripts are expected to include this text:

KY led the observation planning, data analysis, and manuscript preparation. ONC-T operations were conducted by KY, MY, TM, YC, YY, MH, ET, MM, NS, SK, ST, and SSu, under the supervision of TK. YM and SSh managed the planning of spacecraft attitude during the observations. KE and AA provided valuable feedback and contributed to improving the manuscript.
%%WWM came up with the initial research concept and edited the manuscript.
%%OTS obtained the funding and edited the manuscript.
%%EBF provided the formal analysis and validation. He also edited the manuscript.
%%GEH Supervised the undergraduates, wrote the software and administers the project github and Zenodo repositories.
%%
%% Authors can use the Contributor Role Taxonomy (CRediT) at
%% for ideas on how write a good statement tailored to their needs.

\end{contribution}

\clearpage

\appendix

\begin{figure}[h]
    \section{Posterior distributions of
    \texorpdfstring{$T_C$}{TC} and
    \texorpdfstring{$R_p/R_\ast$}{Rp/R*}}
    \label{Appendix:Posterior}
    \centering
    \includegraphics[width=0.8\hsize]{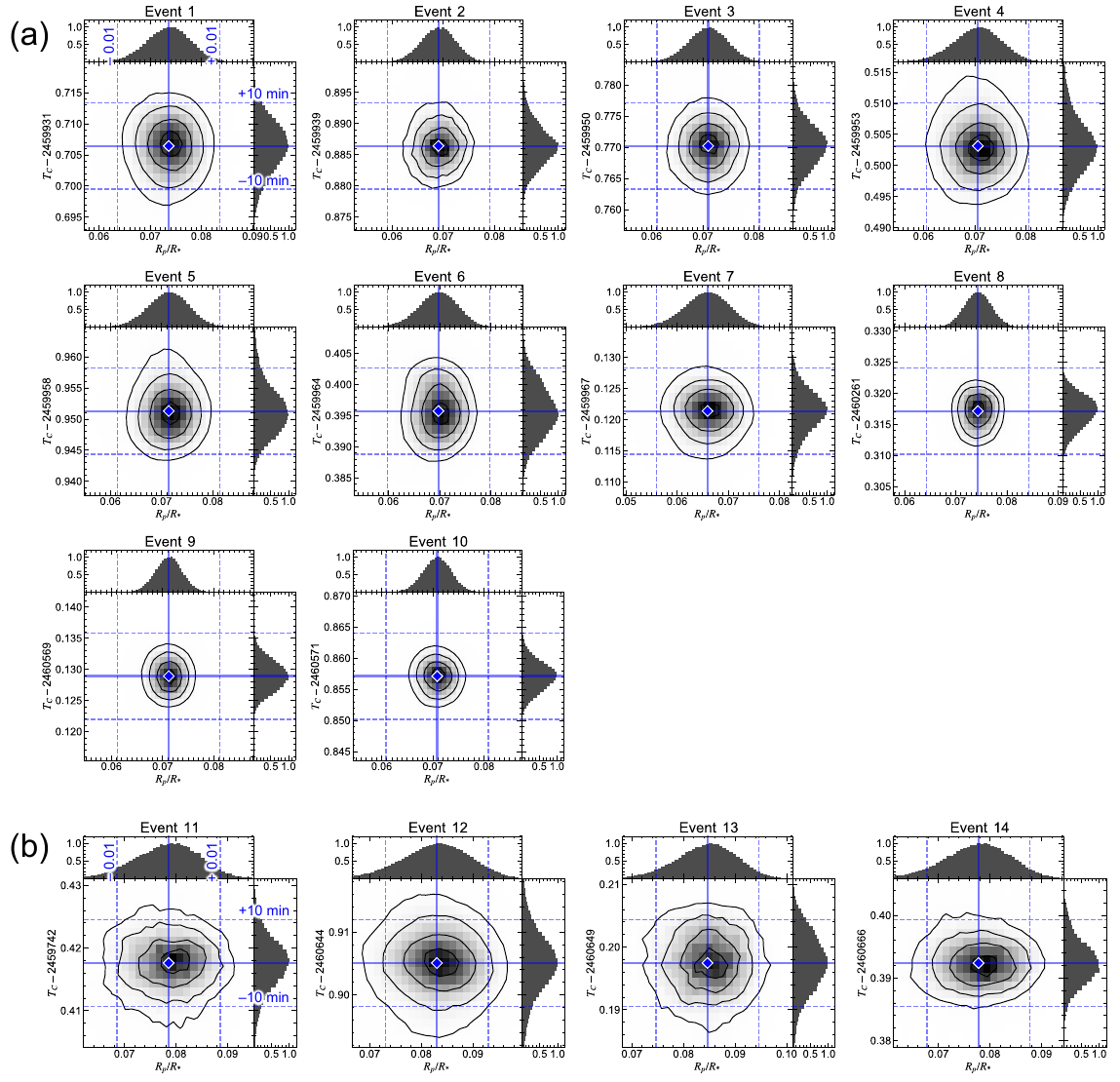}
    \caption{
    Posterior distributions of transit mid-times ($T_C$) and planet-to-star radius ratios ($R_p/R_*$) for each transit of \textbf{(a)} WASP-189 b (Event 1–10) and \textbf{(b)} MASCARA-1 b (Event 11–14) observed by ONC-T. The 2D histograms show the covariance between $R_p/R_*$ and $T_C$, with contours marking the 0.5$\sigma$, 1$\sigma$, 1.5$\sigma$, and 2$\sigma$ confidence levels. The top and right panels display marginal histograms. Blue markers and crosshairs indicate the medians values, while blue dashed lines mark $\pm$0.01 and $\pm$10-minute intervals from the medians of $R_p/R_*$ and $T_C$, respectively.
    }
    \label{Fig_posterior}
\end{figure}

%% For this sample we use BibTeX plus aasjournalv7.bst to generate the
\clearpage
\bibliography{references.bib}

\begin{thebibliography}{}
\expandafter\ifx\csname natexlab\endcsname\relax\def\natexlab#1{#1}\fi
\providecommand{\url}[1]{\href{#1}{#1}}
\providecommand{\dodoi}[1]{doi:~\href{http://doi.org/#1}{\nolinkurl{#1}}}
\providecommand{\doeprint}[1]{\href{http://ascl.net/#1}{\nolinkurl{http://ascl.net/#1}}}
\providecommand{\doarXiv}[1]{\href{https://arxiv.org/abs/#1}{\nolinkurl{https://arxiv.org/abs/#1}}}

\bibitem[{V.~I. {Ananyeva} {et~al.}(2022){Ananyeva}, {Ivanova}, {Shashkova}, {Yakovlev}, {Tavrov}, {Korablev}, \& {Bertaux}}]{Ananyeva_2022}
{Ananyeva}, V.~I., {Ivanova}, A.~E., {Shashkova}, I.~A., {et~al.} 2022, \bibinfo{title}{{The Mass and Orbital-Period Distributions of Exoplanets Accounting for the Observational Selection of the Method for Measuring Radial Velocities. A Dominant (Averaged) Structure of Planetary Systems},} Astronomy Reports, 66, 886, \dodoi{10.1134/S106377292210002X}

\bibitem[{D.~R. {Anderson} {et~al.}(2018){Anderson}, {Temple}, {Nielsen}, {Burdanov}, {Hellier}, {Bouchy}, {Brown}, {Collier Cameron}, {Gillon}, {Jehin}, {Maxted}, {Pepe}, {Pollacco}, {Pozuelos}, {Queloz}, {S{\'e}gransan}, {Smalley}, {Triaud}, {Turner}, {Udry}, \& {West}}]{Anderson_2018}
{Anderson}, D.~R., {Temple}, L.~Y., {Nielsen}, L.~D., {et~al.} 2018, \bibinfo{title}{{WASP-189b: an ultra-hot Jupiter transiting the bright A star HR 5599 in a polar orbit},} arXiv e-prints, arXiv:1809.04897, \dodoi{10.48550/arXiv.1809.04897}

\bibitem[{P. {Chaturvedi} {et~al.}(2014){Chaturvedi}, {Deshpande}, {Dixit}, {Roy}, {Chakraborty}, {Mahadevan}, {Anandarao}, {Hebb}, \& {Janardhan}}]{Chaturvedi_2014}
{Chaturvedi}, P., {Deshpande}, R., {Dixit}, V., {et~al.} 2014, \bibinfo{title}{{Determination of mass and orbital parameters of a low-mass star HD 213597B},} \mnras, 442, 3737, \dodoi{10.1093/mnras/stu1127}

\bibitem[{A. {Cumming} {et~al.}(2008){Cumming}, {Butler}, {Marcy}, {Vogt}, {Wright}, \& {Fischer}}]{Cumming_2008}
{Cumming}, A., {Butler}, R.~P., {Marcy}, G.~W., {et~al.} 2008, \bibinfo{title}{{The Keck Planet Search: Detectability and the Minimum Mass and Orbital Period Distribution of Extrasolar Planets},} \pasp, 120, 531, \dodoi{10.1086/588487}

\bibitem[{D. {Deming} {et~al.}(2015){Deming}, {Knutson}, {Kammer}, {Fulton}, {Ingalls}, {Carey}, {Burrows}, {Fortney}, {Todorov}, {Agol}, {Cowan}, {Desert}, {Fraine}, {Langton}, {Morley}, \& {Showman}}]{Deming_2015}
{Deming}, D., {Knutson}, H., {Kammer}, J., {et~al.} 2015, \bibinfo{title}{{Spitzer Secondary Eclipses of the Dense, Modestly-irradiated, Giant Exoplanet HAT-P-20b Using Pixel-level Decorrelation},} \apj, 805, 132, \dodoi{10.1088/0004-637X/805/2/132}

\bibitem[{C.~D. {Dressing} \& D. {Charbonneau}(2015){Dressing} \& {Charbonneau}}]{Dressing_2015}
{Dressing}, C.~D., \& {Charbonneau}, D. 2015, \bibinfo{title}{{The Occurrence of Potentially Habitable Planets Orbiting M Dwarfs Estimated from the Full Kepler Dataset and an Empirical Measurement of the Detection Sensitivity},} \apj, 807, 45, \dodoi{10.1088/0004-637X/807/1/45}

\bibitem[{A. {Egan} {et~al.}(2023){Egan}, {Nell}, {Suresh}, {France}, {Fleming}, {Sreejith}, {Lambert}, \& {DeCicco}}]{Egan_2023}
{Egan}, A., {Nell}, N., {Suresh}, A., {et~al.} 2023, \bibinfo{title}{{The On-orbit Performance of the Colorado Ultraviolet Transit Experiment Mission},} \aj, 165, 64, \dodoi{10.3847/1538-3881/aca8a3}

\bibitem[{C.~J. {Eyles} {et~al.}(2009){Eyles}, {Harrison}, {Davis}, {Waltham}, {Shaughnessy}, {Mapson-Menard}, {Bewsher}, {Crothers}, {Davies}, {Simnett}, {Howard}, {Moses}, {Newmark}, {Socker}, {Halain}, {Defise}, {Mazy}, \& {Rochus}}]{Eyles_2009}
{Eyles}, C.~J., {Harrison}, R.~A., {Davis}, C.~J., {et~al.} 2009, \bibinfo{title}{{The Heliospheric Imagers Onboard the STEREO Mission},} \solphys, 254, 387, \dodoi{10.1007/s11207-008-9299-0}

\bibitem[{D. {Foreman-Mackey} {et~al.}(2013){Foreman-Mackey}, {Hogg}, {Lang}, \& {Goodman}}]{Foreman-Mackey_2013}
{Foreman-Mackey}, D., {Hogg}, D.~W., {Lang}, D., \& {Goodman}, J. 2013, \bibinfo{title}{{emcee: The MCMC Hammer},} \pasp, 125, 306, \dodoi{10.1086/670067}

\bibitem[{D. {Foreman-Mackey} {et~al.}(2016){Foreman-Mackey}, {Morton}, {Hogg}, {Agol}, \& {Sch{\"o}lkopf}}]{Foreman-Mackey_2016}
{Foreman-Mackey}, D., {Morton}, T.~D., {Hogg}, D.~W., {Agol}, E., \& {Sch{\"o}lkopf}, B. 2016, \bibinfo{title}{{The Population of Long-period Transiting Exoplanets},} \aj, 152, 206, \dodoi{10.3847/0004-6256/152/6/206}

\bibitem[{K. {France} {et~al.}(2023){France}, {Fleming}, {Egan}, {Desert}, {Fossati}, {Koskinen}, {Nell}, {Petit}, {Vidotto}, {Beasley}, {DeCicco}, {Sreejith}, {Suresh}, {Baumert}, {Cauley}, {Villarreal D'Angelo}, {Hoadley}, {Kane}, {Kohnert}, {Lambert}, \& {Ulrich}}]{France_2023}
{France}, K., {Fleming}, B., {Egan}, A., {et~al.} 2023, \bibinfo{title}{{The Colorado Ultraviolet Transit Experiment Mission Overview},} \aj, 165, 63, \dodoi{10.3847/1538-3881/aca8a2}

\bibitem[{A. {Fuentes} \& M. {Solar}(2024){Fuentes} \& {Solar}}]{Fuentes_2024}
{Fuentes}, A., \& {Solar}, M. 2024, \bibinfo{title}{{Synthetic light curves of exoplanet transit using nanosatellite data},} Astronomy and Computing, 47, 100816, \dodoi{10.1016/j.ascom.2024.100816}

\bibitem[{S. Grocott {et~al.}(2004)Grocott, Zee, \& Matthews}]{Grocott_2004}
Grocott, S., Zee, R., \& Matthews, J. 2004, in Proceedings of the Small Satellite Conference (SmallSat) 2004.
\newblock \url{https://digitalcommons.usu.edu/smallsat/2004/All2004/50/}

\bibitem[{M. {Hippke} \& R. {Heller}(2019){Hippke} \& {Heller}}]{Hippke_2019}
{Hippke}, M., \& {Heller}, R. 2019, \bibinfo{title}{{Optimized transit detection algorithm to search for periodic transits of small planets},} \aap, 623, A39, \dodoi{10.1051/0004-6361/201834672}

\bibitem[{E. {H{\o}g} {et~al.}(2000){H{\o}g}, {Fabricius}, {Makarov}, {Urban}, {Corbin}, {Wycoff}, {Bastian}, {Schwekendiek}, \& {Wicenec}}]{Hog_2000}
{H{\o}g}, E., {Fabricius}, C., {Makarov}, V.~V., {et~al.} 2000, \bibinfo{title}{{The Tycho-2 catalogue of the 2.5 million brightest stars},} \aap, 355, L27

\bibitem[{M.~J. {Hooton} {et~al.}(2022){Hooton}, {Hoyer}, {Kitzmann}, {Morris}, {Smith}, {Collier Cameron}, {Futyan}, {Maxted}, {Queloz}, {Demory}, {Heng}, {Lendl}, {Cabrera}, {Csizmadia}, {Deline}, {Parviainen}, {Salmon}, {Sulis}, {Wilson}, {Bonfanti}, {Brandeker}, {Demangeon}, {Oshagh}, {Persson}, {Scandariato}, {Alibert}, {Alonso}, {Anglada Escud{\'e}}, {B{\'a}rczy}, {Barrado}, {Barros}, {Baumjohann}, {Beck}, {Beck}, {Benz}, {Billot}, {Bonfils}, {Bourrier}, {Broeg}, {Busch}, {Charnoz}, {Davies}, {Deleuil}, {Delrez}, {Ehrenreich}, {Erikson}, {Farinato}, {Fortier}, {Fossati}, {Fridlund}, {Gandolfi}, {Gillon}, {G{\"u}del}, {Isaak}, {Jones}, {Kiss}, {Laskar}, {Lecavelier des Etangs}, {Lovis}, {Luntzer}, {Magrin}, {Nascimbeni}, {Olofsson}, {Ottensamer}, {Pagano}, {Pall{\'e}}, {Peter}, {Piotto}, {Pollacco}, {Ragazzoni}, {Rando}, {Ratti}, {Rauer}, {Ribas}, {Santos}, {S{\'e}gransan}, {Simon}, {Sousa}, {Steller}, {Szab{\'o}}, {Thomas}, {Udry}, {Ulmer}, {Van Grootel}, \& {Walton}}]{Hooton_2022}
{Hooton}, M.~J., {Hoyer}, S., {Kitzmann}, D., {et~al.} 2022, \bibinfo{title}{{Spi-OPS: Spitzer and CHEOPS confirm the near-polar orbit of MASCARA-1 b and reveal a hint of dayside reflection},} \aap, 658, A75, \dodoi{10.1051/0004-6361/202141645}

\bibitem[{E.~S. {Ivshina} \& J.~N. {Winn}(2022){Ivshina} \& {Winn}}]{Ivshina_2022}
{Ivshina}, E.~S., \& {Winn}, J.~N. 2022, \bibinfo{title}{{TESS Transit Timing of Hundreds of Hot Jupiters},} \apjs, 259, 62, \dodoi{10.3847/1538-4365/ac545b}

\bibitem[{J.~M. {Jenkins} {et~al.}(2016){Jenkins}, {Twicken}, {McCauliff}, {Campbell}, {Sanderfer}, {Lung}, {Mansouri-Samani}, {Girouard}, {Tenenbaum}, {Klaus}, {Smith}, {Caldwell}, {Chacon}, {Henze}, {Heiges}, {Latham}, {Morgan}, {Swade}, {Rinehart}, \& {Vanderspek}}]{Jenkins_2016}
{Jenkins}, J.~M., {Twicken}, J.~D., {McCauliff}, S., {et~al.} 2016, in Society of Photo-Optical Instrumentation Engineers (SPIE) Conference Series, Vol. 9913, Software and Cyberinfrastructure for Astronomy IV, ed. G.~{Chiozzi} \& J.~C. {Guzman}, 99133E, \dodoi{10.1117/12.2233418}

\bibitem[{S. {Kameda} {et~al.}(2017){Kameda}, {Suzuki}, {Takamatsu}, {Cho}, {Yasuda}, {Yamada}, {Sawada}, {Honda}, {Morota}, {Honda}, {Sato}, {Okumura}, {Shibasaki}, {Ikezawa}, \& {Sugita}}]{Kameda_2017}
{Kameda}, S., {Suzuki}, H., {Takamatsu}, T., {et~al.} 2017, \bibinfo{title}{{Preflight Calibration Test Results for Optical Navigation Camera Telescope (ONC-T) Onboard the Hayabusa2 Spacecraft},} \ssr, 208, 17, \dodoi{10.1007/s11214-015-0227-y}

\bibitem[{S. Kameda {et~al.}(2021)Kameda, Yokota, Kouyama, Tatsumi, Ishida, Morota, Honda, Sakatani, Yamada, Matsuoka, Suzuki, Cho, Hayakawa, Honda, Sawada, Yoshioka, Ogawa, \& Sugita}]{Kameda_2021}
Kameda, S., Yokota, Y., Kouyama, T., {et~al.} 2021, \bibinfo{title}{Improved method of hydrous mineral detection by latitudinal distribution of 0.7-μm surface reflectance absorption on the asteroid Ryugu,} Icarus, 360, 114348, \dodoi{https://doi.org/10.1016/j.icarus.2021.114348}

\bibitem[{H. {Kawahara} {et~al.}(2020){Kawahara}, {Masuda}, {Kotani}, {Tada}, {Kataza}, {Ikari}, {Aohama}, {Hosonuma}, {Mikuriya}, {Ikoma}, {Kasahara}, {Sako}, {Sugita}, {Tatsumi}, \& {Yoshioka}}]{Kawahara_2020}
{Kawahara}, H., {Masuda}, K., {Kotani}, T., {et~al.} 2020, in Society of Photo-Optical Instrumentation Engineers (SPIE) Conference Series, Vol. 11443, Space Telescopes and Instrumentation 2020: Optical, Infrared, and Millimeter Wave, ed. M.~{Lystrup} \& M.~D. {Perrin}, 1144316, \dodoi{10.1117/12.2562266}

\bibitem[{M. {Knapp} {et~al.}(2020){Knapp}, {Seager}, {Demory}, {Krishnamurthy}, {Smith}, {Pong}, {Bailey}, {Donner}, {Pasquale}, {Campuzano}, {Smith}, {Luu}, {Babuscia}, {Bocchino}, {Loveland}, {Colley}, {Gedenk}, {Kulkarni}, {Hughes}, {White}, {Krajewski}, \& {Fesq}}]{Knapp_2020}
{Knapp}, M., {Seager}, S., {Demory}, B.-O., {et~al.} 2020, \bibinfo{title}{{Demonstrating High-precision Photometry with a CubeSat: ASTERIA Observations of 55 Cancri e},} \aj, 160, 23, \dodoi{10.3847/1538-3881/ab8bcc}

\bibitem[{A. {Kokori} {et~al.}(2022){Kokori}, {Tsiaras}, {Edwards}, {Rocchetto}, {Tinetti}, {Bewersdorff}, {Jongen}, {Lekkas}, {Pantelidou}, {Poultourtzidis}, {W{\"u}nsche}, {Aggelis}, {Agnihotri}, {Arena}, {Bachschmidt}, {Bennett}, {Benni}, {Bernacki}, {Besson}, {Betti}, {Biagini}, {Brandebourg}, {Bretton}, {Brincat}, {Cal{\'o}}, {Campos}, {Casali}, {Ciantini}, {Crow}, {Dauchet}, {Dawes}, {Deldem}, {Deligeorgopoulos}, {Dymock}, {Eenm{\"a}e}, {Evans}, {Esseiva}, {Falco}, {Ferratfiat}, {Fowler}, {Futcher}, {Gaitan}, {Horta}, {Guerra}, {Hurter}, {Jones}, {Kang}, {Kiiskinen}, {Kim}, {Laloum}, {Lee}, {Lomoz}, {Lopresti}, {Mallonn}, {Mannucci}, {Marino}, {Mario}, {Marquette}, {Michelet}, {Miller}, {Mollier}, {Molina}, {Montigiani}, {Mortari}, {Morvan}, {Mugnai}, {Naponiello}, {Nastasi}, {Neito}, {Pace}, {Papadeas}, {Paschalis}, {Pereira}, {Perroud}, {Phillips}, {Pintr}, {Pioppa}, {Popowicz}, {Raetz}, {Regembal}, {Rickard}, {Roberts}, {Rousselot}, {Rubia}, {Savage}, {Sedita}, {Shave-Wall}, {Sioulas},
  {{\v{S}}koln{\'\i}k}, {Smith}, {St-Gelais}, {Stouraitis}, {Strikis}, {Thurston}, {Tomacelli}, {Tomatis}, {Trevan}, {Valeau}, {Vignes}, {Vora}, {Vra{\v{s}}{\v{t}}{\'a}k}, {Walter}, {Wenzel}, {Wright}, \& {Z{\'\i}bar}}]{Kokori_2022}
{Kokori}, A., {Tsiaras}, A., {Edwards}, B., {et~al.} 2022, \bibinfo{title}{{ExoClock Project. II. A Large-scale Integrated Study with 180 Updated Exoplanet Ephemerides},} \apjs, 258, 40, \dodoi{10.3847/1538-4365/ac3a10}

\bibitem[{A. {Kokori} {et~al.}(2023){Kokori}, {Tsiaras}, {Edwards}, {Jones}, {Pantelidou}, {Tinetti}, {Bewersdorff}, {Iliadou}, {Jongen}, {Lekkas}, {Nastasi}, {Poultourtzidis}, {Sidiropoulos}, {Walter}, {W{\"u}nsche}, {Abraham}, {Agnihotri}, {Albanesi}, {Arce-Mansego}, {Arnot}, {Audejean}, {Aumasson}, {Bachschmidt}, {Baj}, {Barroy}, {Belinski}, {Bennett}, {Benni}, {Bernacki}, {Betti}, {Biagini}, {Bosch}, {Brandebourg}, {Br{\'a}t}, {Bretton}, {Brincat}, {Brouillard}, {Bruzas}, {Bruzzone}, {Buckland}, {Cal{\'o}}, {Campos}, {Carre{\~n}o}, {Carrion Rodrigo}, {Casali}, {Casalnuovo}, {Cataneo}, {Chang}, {Changeat}, {Chowdhury}, {Ciantini}, {Cilluffo}, {Coliac}, {Conzo}, {Correa}, {Coulon}, {Crouzet}, {Crow}, {Curtis}, {Daniel}, {Dauchet}, {Dawes}, {Deldem}, {Deligeorgopoulos}, {Dransfield}, {Dymock}, {Eenm{\"a}e}, {Esseiva}, {Evans}, {Falco}, {Farf{\'a}n}, {Fern{\'a}ndez-Laj{\'u}s}, {Ferratfiat}, {Ferreira}, {Ferretti}, {Fio{\l}ka}, {Fowler}, {Futcher}, {Gabellini}, {Gainey}, {Gaitan}, {Gajdo{\v{s}}},
  {Garc{\'\i}a-S{\'a}nchez}, {Garlitz}, {Gillier}, {Gison}, {Gonzales}, {Gorshanov}, {Grau Horta}, {Grivas}, {Guerra}, {Guillot}, {Haswell}, {Haymes}, {Hentunen}, {Hills}, {Hose}, {Humbert}, {Hurter}, {Hynek}, {Irzyk}, {Jacobsen}, {Jannetta}, {Johnson}, {J{\'o}{\'z}wik-Wabik}, {Kaeouach}, {Kang}, {Kiiskinen}, {Kim}, {Kivila}, {Koch}, {Kolb}, {Ku{\v{c}}{\'a}kov{\'a}}, {Lai}, {Laloum}, {Lasota}, {Lewis}, {Liakos}, {Libotte}, {Lomoz}, {Lopresti}, {Majewski}, {Malcher}, {Mallonn}, {Mannucci}, {Marchini}, {Mari}, {Marino}, {Marino}, {Mario}, {Marquette}, {Mart{\'\i}nez-Bravo}, {Ma{\v{s}}ek}, {Matassa}, {Michel}, {Michelet}, {Miller}, {Miny}, {Molina}, {Mollier}, {Monteleone}, {Montigiani}, {Morales-Aimar}, {Mortari}, {Morvan}, {Mugnai}, {Murawski}, {Naponiello}, {Naudin}, {Naves}, {N{\'e}el}, {Neito}, {Neveu}, {Noschese}, {{\"O}{\u{g}}men}, {Ohshima}, {Orbanic}, {Pace}, {Pantacchini}, {Paschalis}, {Pereira}, {Peretto}, {Perroud}, {Phillips}, {Pintr}, {Pioppa}, {Plazas}, {Poelarends}, {Popowicz}, {Purcell},
  {Quinn}, {Raetz}, {Rees}, {Regembal}, {Rocchetto}, {Rocci}, {Rockenbauer}, {Roth}, {Rousselot}, {Rubia}, {Ruocco}, {Russo}, {Salisbury}, {Salvaggio}, {Santos}, {Savage}, {Scaggiante}, {Sedita}, {Shadick}, {Silva}, {Sioulas}, {{\v{S}}koln{\'\i}k}, {Smith}, {Smolka}, {Solmaz}, {Stanbury}, {Stouraitis}, {Tan}, {Theusner}, \& {Thurston}}]{Kokori_2023}
{Kokori}, A., {Tsiaras}, A., {Edwards}, B., {et~al.} 2023, \bibinfo{title}{{ExoClock Project. III. 450 New Exoplanet Ephemerides from Ground and Space Observations},} \apjs, 265, 4, \dodoi{10.3847/1538-4365/ac9da4}

\bibitem[{T. {Kouyama} {et~al.}(2021){Kouyama}, {Tatsumi}, {Yokota}, {Yumoto}, {Yamada}, {Honda}, {Kameda}, {Suzuki}, {Sakatani}, {Hayakawa}, {Morota}, {Matsuoka}, {Cho}, {Honda}, {Sawada}, {Yoshioka}, \& {Sugita}}]{Kouyama_2021}
{Kouyama}, T., {Tatsumi}, E., {Yokota}, Y., {et~al.} 2021, \bibinfo{title}{{Post-arrival calibration of Hayabusa2's optical navigation cameras (ONCs): Severe effects from touchdown events},} \icarus, 360, 114353, \dodoi{10.1016/j.icarus.2021.114353}

\bibitem[{A. {Krishnamurthy} {et~al.}(2021){Krishnamurthy}, {Knapp}, {G{\"u}nther}, {Daylan}, {Demory}, {Seager}, {Bailey}, {Smith}, {Pong}, {Hughes}, {Donner}, {Di Pasquale}, {Campuzano}, {Smith}, {Luu}, {Babuscia}, {Bocchino}, {Loveland}, {Colley}, {Gedenk}, {Kulkarni}, {White}, {Krajewski}, \& {Fesq}}]{Krishnamurthy_2021}
{Krishnamurthy}, A., {Knapp}, M., {G{\"u}nther}, M.~N., {et~al.} 2021, \bibinfo{title}{{Transit Search for Exoplanets around Alpha Centauri A and B with ASTERIA},} \aj, 161, 275, \dodoi{10.3847/1538-3881/abf2c0}

\bibitem[{V. Lapeyrere {et~al.}(2017)Lapeyrere, Lacour, David, Nowak, Crouzier, Schworer, Perrot, \& Rayane}]{Lapeyrere_2017}
Lapeyrere, V., Lacour, S., David, L., {et~al.} 2017, in Proceedings of the 31st Annual AIAA/USU Conference on Small Satellites.
\newblock \url{https://digitalcommons.usu.edu/smallsat/2017/all2017/86/}

\bibitem[{M. {Lendl} {et~al.}(2020){Lendl}, {Csizmadia}, {Deline}, {Fossati}, {Kitzmann}, {Heng}, {Hoyer}, {Salmon}, {Benz}, {Broeg}, {Ehrenreich}, {Fortier}, {Queloz}, {Bonfanti}, {Brandeker}, {Collier Cameron}, {Delrez}, {Garcia Mu{\~n}oz}, {Hooton}, {Maxted}, {Morris}, {Van Grootel}, {Wilson}, {Alibert}, {Alonso}, {Asquier}, {Bandy}, {B{\'a}rczy}, {Barrado}, {Barros}, {Baumjohann}, {Beck}, {Beck}, {Bekkelien}, {Bergomi}, {Billot}, {Biondi}, {Bonfils}, {Bourrier}, {Busch}, {Cabrera}, {Cessa}, {Charnoz}, {Chazelas}, {Corral Van Damme}, {Davies}, {Deleuil}, {Demangeon}, {Demory}, {Erikson}, {Farinato}, {Fridlund}, {Futyan}, {Gandolfi}, {Gillon}, {Guterman}, {Hasiba}, {Hernandez}, {Isaak}, {Kiss}, {Kuntzer}, {Lecavelier des Etangs}, {L{\"u}ftinger}, {Laskar}, {Lovis}, {Magrin}, {Malvasio}, {Marafatto}, {Michaelis}, {Munari}, {Nascimbeni}, {Olofsson}, {Ottacher}, {Ottensamer}, {Pagano}, {Pall{\'e}}, {Peter}, {Piazza}, {Piotto}, {Pollacco}, {Ratti}, {Rauer}, {Ragazzoni}, {Rando}, {Ribas}, {Rieder}, {Rohlfs},
  {Safa}, {Santos}, {Scandariato}, {S{\'e}gransan}, {Simon}, {Singh}, {Smith}, {Sordet}, {Sousa}, {Steller}, {Szab{\'o}}, {Thomas}, {Tschentscher}, {Udry}, {Viotto}, {Walter}, {Walton}, {Wildi}, \& {Wolter}}]{Lendl_2020}
{Lendl}, M., {Csizmadia}, S., {Deline}, A., {et~al.} 2020, \bibinfo{title}{{The hot dayside and asymmetric transit of WASP-189 b seen by CHEOPS},} \aap, 643, A94, \dodoi{10.1051/0004-6361/202038677}

\bibitem[{J. {Lillo-Box} {et~al.}(2020){Lillo-Box}, {Figueira}, {Leleu}, {Acu{\~n}a}, {Faria}, {Hara}, {Santos}, {Correia}, {Robutel}, {Deleuil}, {Barrado}, {Sousa}, {Bonfils}, {Mousis}, {Almenara}, {Astudillo-Defru}, {Marcq}, {Udry}, {Lovis}, \& {Pepe}}]{Lillo-Box_2020}
{Lillo-Box}, J., {Figueira}, P., {Leleu}, A., {et~al.} 2020, \bibinfo{title}{{Planetary system LHS 1140 revisited with ESPRESSO and TESS},} \aap, 642, A121, \dodoi{10.1051/0004-6361/202038922}

\bibitem[{K. {Mandel} \& E. {Agol}(2002){Mandel} \& {Agol}}]{Mandel_Agol_2002}
{Mandel}, K., \& {Agol}, E. 2002, \bibinfo{title}{{Analytic Light Curves for Planetary Transit Searches},} \apjl, 580, L171, \dodoi{10.1086/345520}

\bibitem[{R.~M. {Millan} {et~al.}(2019){Millan}, {von Steiger}, {Ariel}, {Bartalev}, {Borgeaud}, {Campagnola}, {Castillo-Rogez}, {Fl{\'e}ron}, {Gass}, {Gregorio}, {Klumpar}, {Lal}, {Macdonald}, {Park}, {Sambasiva Rao}, {Schilling}, {Stephens}, {Title}, \& {Wu}}]{Millan_2019}
{Millan}, R.~M., {von Steiger}, R., {Ariel}, M., {et~al.} 2019, \bibinfo{title}{{Small satellites for space science. A COSPAR scientific roadmap},} Advances in Space Research, 64, 1466, \dodoi{10.1016/j.asr.2019.07.035}

\bibitem[{E. {Miller-Ricci} {et~al.}(2008){Miller-Ricci}, {Rowe}, {Sasselov}, {Matthews}, {Kuschnig}, {Croll}, {Guenther}, {Moffat}, {Rucinski}, {Walker}, \& {Weiss}}]{Miller-Ricci_2008}
{Miller-Ricci}, E., {Rowe}, J.~F., {Sasselov}, D., {et~al.} 2008, \bibinfo{title}{{MOST Space-based Photometry of the Transiting Exoplanet System HD 189733: Precise Timing Measurements for Transits across an Active Star},} \apj, 682, 593, \dodoi{10.1086/587634}

\bibitem[{M. {Mol Lous} {et~al.}(2018){Mol Lous}, {Weenk}, {Kenworthy}, {Zwintz}, \& {Kuschnig}}]{Lous_2018}
{Mol Lous}, M., {Weenk}, E., {Kenworthy}, M.~A., {Zwintz}, K., \& {Kuschnig}, R. 2018, \bibinfo{title}{{A search for transiting planets in the {\ensuremath{\beta}} Pictoris system},} \aap, 615, A145, \dodoi{10.1051/0004-6361/201731941}

\bibitem[{ {National Academies of Sciences, Engineering, and Medicine}(2018){National Academies of Sciences, Engineering, and Medicine}}]{NationalAcademies_2018}
{National Academies of Sciences, Engineering, and Medicine}. 2018, \bibinfo{title}{Exoplanet Science Strategy,} Washington, DC: The National Academies Press, \dodoi{10.17226/25187}

\bibitem[{T. {Nguyen} {et~al.}(2018){Nguyen}, {Morgan}, {Vanderspek}, {Levine}, {Kephart}, {Francis}, {Zapetis}, {Cahoy}, \& {Ricker}}]{Nguyen_2018}
{Nguyen}, T., {Morgan}, E., {Vanderspek}, R., {et~al.} 2018, \bibinfo{title}{{Fine-pointing performance and corresponding photometric precision of the Transiting Exoplanet Survey Satellite},} Journal of Astronomical Telescopes, Instruments, and Systems, 4, 047001, \dodoi{10.1117/1.JATIS.4.4.047001}

\bibitem[{M. {Nowak} {et~al.}(2018){Nowak}, {Lacour}, {Crouzier}, {David}, {Lapeyr{\`e}re}, \& {Schworer}}]{Nowak_2018}
{Nowak}, M., {Lacour}, S., {Crouzier}, A., {et~al.} 2018, in Society of Photo-Optical Instrumentation Engineers (SPIE) Conference Series, Vol. 10698, Space Telescopes and Instrumentation 2018: Optical, Infrared, and Millimeter Wave, ed. M.~{Lystrup}, H.~A. {MacEwen}, G.~G. {Fazio}, N.~{Batalha}, N.~{Siegler}, \& E.~C. {Tong}, 1069821, \dodoi{10.1117/12.2313242}

\bibitem[{M. {Nowak} {et~al.}(2017){Nowak}, {Lacour}, {Lapeyr{\`e}re}, {David}, {Crouzier}, {Schworer}, {Perrot}, \& {Rayane}}]{Nowak_2017}
{Nowak}, M., {Lacour}, S., {Lapeyr{\`e}re}, V., {et~al.} 2017, \bibinfo{title}{{A Compact and Lightweight Fibered Photometer for the PicSat Mission},} arXiv e-prints, arXiv:1708.04015, \dodoi{10.48550/arXiv.1708.04015}

\bibitem[{N. {Ogawa} {et~al.}(2020){Ogawa}, {Terui}, {Mimasu}, {Yoshikawa}, {Ono}, {Yasuda}, {Matsushima}, {Masuda}, {Hihara}, {Sano}, {Matsuhisa}, {Danno}, {Yamada}, {Yokota}, {Takei}, {Saiki}, \& {Tsuda}}]{Ogawa_2020}
{Ogawa}, N., {Terui}, F., {Mimasu}, Y., {et~al.} 2020, \bibinfo{title}{{Image-based autonomous navigation of Hayabusa2 using artificial landmarks: The design and brief in-flight results of the first landing on asteroid Ryugu},} Astrodynamics, 4, 89, \dodoi{10.1007/s42064-020-0070-0}

\bibitem[{H. {Parviainen}(2015){Parviainen}}]{Parviainen_2015}
{Parviainen}, H. 2015, \bibinfo{title}{{PYTRANSIT: fast and easy exoplanet transit modelling in PYTHON},} \mnras, 450, 3233, \dodoi{10.1093/mnras/stv894}

\bibitem[{K.~C. {Patra} {et~al.}(2017){Patra}, {Winn}, {Holman}, {Yu}, {Deming}, \& {Dai}}]{Patra_2017}
{Patra}, K.~C., {Winn}, J.~N., {Holman}, M.~J., {et~al.} 2017, \bibinfo{title}{{The Apparently Decaying Orbit of WASP-12b},} \aj, 154, 4, \dodoi{10.3847/1538-3881/aa6d75}

\bibitem[{A. {Popowicz}(2018){Popowicz}}]{Popowicz_2018}
{Popowicz}, A. 2018, \bibinfo{title}{{Analysis of Dark Current in BRITE Nanostellite CCD Sensors},} Sensors, 18, 479, \dodoi{10.3390/s18020479}

\bibitem[{G.~R. {Ricker} {et~al.}(2015){Ricker}, {Winn}, {Vanderspek}, {Latham}, {Bakos}, {Bean}, {Berta-Thompson}, {Brown}, {Buchhave}, {Butler}, {Butler}, {Chaplin}, {Charbonneau}, {Christensen-Dalsgaard}, {Clampin}, {Deming}, {Doty}, {De Lee}, {Dressing}, {Dunham}, {Endl}, {Fressin}, {Ge}, {Henning}, {Holman}, {Howard}, {Ida}, {Jenkins}, {Jernigan}, {Johnson}, {Kaltenegger}, {Kawai}, {Kjeldsen}, {Laughlin}, {Levine}, {Lin}, {Lissauer}, {MacQueen}, {Marcy}, {McCullough}, {Morton}, {Narita}, {Paegert}, {Palle}, {Pepe}, {Pepper}, {Quirrenbach}, {Rinehart}, {Sasselov}, {Sato}, {Seager}, {Sozzetti}, {Stassun}, {Sullivan}, {Szentgyorgyi}, {Torres}, {Udry}, \& {Villasenor}}]{Ricker_2015}
{Ricker}, G.~R., {Winn}, J.~N., {Vanderspek}, R., {et~al.} 2015, \bibinfo{title}{{Transiting Exoplanet Survey Satellite (TESS)},} Journal of Astronomical Telescopes, Instruments, and Systems, 1, 014003, \dodoi{10.1117/1.JATIS.1.1.014003}

\bibitem[{K. Sarda {et~al.}(2014)Sarda, Grant, Chaumont, Choi, Johnston-Lemke, \& Zee}]{Sarda_2014}
Sarda, K., Grant, C., Chaumont, M., {et~al.} 2014, in Proceedings of the 28th Annual AIAA/USU Conference on Small Satellites (SmallSat).
\newblock \url{https://digitalcommons.usu.edu/smallsat/2014/YearReview/4/}

\bibitem[{S. {Serjeant} {et~al.}(2020){Serjeant}, {Elvis}, \& {Tinetti}}]{Serjeant_2020}
{Serjeant}, S., {Elvis}, M., \& {Tinetti}, G. 2020, \bibinfo{title}{{The future of astronomy with small satellites},} Nature Astronomy, 4, 1031, \dodoi{10.1038/s41550-020-1201-5}

\bibitem[{E.~L. {Shkolnik}(2018){Shkolnik}}]{Shkolnik_2018}
{Shkolnik}, E.~L. 2018, \bibinfo{title}{{On the verge of an astronomy CubeSat revolution},} Nature Astronomy, 2, 374, \dodoi{10.1038/s41550-018-0438-8}

\bibitem[{M. Smith {et~al.}(2018)Smith, Donner, Knapp, Pong, Smith, Luu, Di~Pasquale, \& Campuzano}]{Smith_2018}
Smith, M., Donner, A., Knapp, M., {et~al.} 2018, in Proceedings of the 32nd Annual AIAA/USU Conference on Small Satellites.
\newblock \url{https://digitalcommons.usu.edu/smallsat/2018/all2018/255/}

\bibitem[{A.~G. {Sreejith} {et~al.}(2023){Sreejith}, {France}, {Fossati}, {Koskinen}, {Egan}, {Cauley}, {Cubillos}, {Ambily}, {Huang}, {Lavvas}, {Fleming}, {Desert}, {Nell}, {Petit}, \& {Vidotto}}]{Sreejith_2023}
{Sreejith}, A.~G., {France}, K., {Fossati}, L., {et~al.} 2023, \bibinfo{title}{{CUTE Reveals Escaping Metals in the Upper Atmosphere of the Ultrahot Jupiter WASP-189b},} \apjl, 954, L23, \dodoi{10.3847/2041-8213/acef1c}

\bibitem[{H. {Suzuki} {et~al.}(2018){Suzuki}, {Yamada}, {Kouyama}, {Tatsumi}, {Kameda}, {Honda}, {Sawada}, {Ogawa}, {Morota}, {Honda}, {Sakatani}, {Hayakawa}, {Yokota}, {Yamamoto}, \& {Sugita}}]{Suzuki_2018}
{Suzuki}, H., {Yamada}, M., {Kouyama}, T., {et~al.} 2018, \bibinfo{title}{{Initial inflight calibration for Hayabusa2 optical navigation camera (ONC) for science observations of asteroid Ryugu},} \icarus, 300, 341, \dodoi{10.1016/j.icarus.2017.09.011}

\bibitem[{G.~J.~J. {Talens} {et~al.}(2017){Talens}, {Albrecht}, {Spronck}, {Lesage}, {Otten}, {Stuik}, {Van Eylen}, {Van Winckel}, {Pollacco}, {McCormac}, {Grundahl}, {Fredslund Andersen}, {Antoci}, \& {Snellen}}]{Talens_2017}
{Talens}, G.~J.~J., {Albrecht}, S., {Spronck}, J.~F.~P., {et~al.} 2017, \bibinfo{title}{{MASCARA-1 b. A hot Jupiter transiting a bright m$_{V}$ = 8.3 A-star in a misaligned orbit},} \aap, 606, A73, \dodoi{10.1051/0004-6361/201731282}

\bibitem[{E. {Tatsumi} {et~al.}(2019){Tatsumi}, {Kouyama}, {Suzuki}, {Yamada}, {Sakatani}, {Kameda}, {Yokota}, {Honda}, {Morota}, {Moroi}, {Tanabe}, {Kamiyoshihara}, {Ishida}, {Yoshioka}, {Sato}, {Honda}, {Hayakawa}, {Kitazato}, {Sawada}, \& {Sugita}}]{Tatsumi_2019}
{Tatsumi}, E., {Kouyama}, T., {Suzuki}, H., {et~al.} 2019, \bibinfo{title}{{Updated inflight calibration of Hayabusa2's optical navigation camera (ONC) for scientific observations during the cruise phase},} \icarus, 325, 153, \dodoi{10.1016/j.icarus.2019.01.015}

\bibitem[{Y. Tsuda {et~al.}(2017)Tsuda, Ono, Saiki, Mimasu, Ogawa, \& Terui}]{Tsuda_2017}
Tsuda, Y., Ono, G., Saiki, T., {et~al.} 2017, \bibinfo{title}{Solar radiation pressure-assisted fuel-free Sun tracking and its application to Hayabusa2,} Journal of Spacecraft and Rockets, 54, 1284

\bibitem[{J.~D. Twicken {et~al.}(2020)Twicken, Caldwell, Jenkins, Vanderspek, Tenenbaum, Smith, Wohler, Rose, Ting, Vanderspek, {et~al.}}]{Twicken_2020}
Twicken, J.~D., Caldwell, D.~A., Jenkins, J.~M., {et~al.} 2020, \bibinfo{title}{Tess science data products description document: Exp-tess-arc-icd-0014 rev f,}

\bibitem[{V. {Van Grootel} {et~al.}(2021){Van Grootel}, {Pozuelos}, {Thuillier}, {Charpinet}, {Delrez}, {Beck}, {Fortier}, {Hoyer}, {Sousa}, {Barlow}, {Billot}, {D{\'e}vora-Pajares}, {{\O}stensen}, {Alibert}, {Alonso}, {Anglada Escud{\'e}}, {Asquier}, {Barrado}, {Barros}, {Baumjohann}, {Beck}, {Bekkelien}, {Benz}, {Bonfils}, {Brandeker}, {Broeg}, {Bruno}, {B{\'a}rczy}, {Cabrera}, {Cameron}, {Charnoz}, {Davies}, {Deleuil}, {Demangeon}, {Demory}, {Ehrenreich}, {Erikson}, {Fossati}, {Fridlund}, {Futyan}, {Gandolfi}, {Gillon}, {Guedel}, {Heng}, {Isaak}, {Kiss}, {Laskar}, {Lecavelier des Etangs}, {Lendl}, {Lovis}, {Magrin}, {Maxted}, {Mecina}, {Mustill}, {Nascimbeni}, {Olofsson}, {Ottensamer}, {Pagano}, {Pall{\'e}}, {Peter}, {Piotto}, {Plesseria}, {Pollacco}, {Queloz}, {Ragazzoni}, {Rando}, {Rauer}, {Ribas}, {Santos}, {Scandariato}, {S{\'e}gransan}, {Silvotti}, {Simon}, {Smith}, {Steller}, {Szab{\'o}}, {Thomas}, {Udry}, {Viotto}, {Walton}, {Westerdorff}, \& {Wilson}}]{VanGrootel_2021}
{Van Grootel}, V., {Pozuelos}, F.~J., {Thuillier}, A., {et~al.} 2021, \bibinfo{title}{{A search for transiting planets around hot subdwarfs. I. Methods and performance tests on light curves from Kepler, K2, TESS, and CHEOPS},} \aap, 650, A205, \dodoi{10.1051/0004-6361/202140381}

\bibitem[{R. Vanderspek {et~al.}(2018)Vanderspek, Doty, Fausnaugh, {et~al.}}]{Vanderspek_2018}
Vanderspek, R., Doty, J., Fausnaugh, M., {et~al.} 2018, \bibinfo{title}{TESS Instrument Handbook, Tech. Rep., Kavli Institute for Astrophysics and Space Science, Massachusetts Institute of Technology,}

\bibitem[{G. {Walker} {et~al.}(2003){Walker}, {Matthews}, {Kuschnig}, {Johnson}, {Rucinski}, {Pazder}, {Burley}, {Walker}, {Skaret}, {Zee}, {Grocott}, {Carroll}, {Sinclair}, {Sturgeon}, \& {Harron}}]{Walker_2003}
{Walker}, G., {Matthews}, J., {Kuschnig}, R., {et~al.} 2003, \bibinfo{title}{{The MOST Asteroseismology Mission: Ultraprecise Photometry from Space},} \pasp, 115, 1023, \dodoi{10.1086/377358}

\bibitem[{W.~W. {Weiss} {et~al.}(2021){Weiss}, {Zwintz}, {Kuschnig}, {Handler}, {Moffat}, {Baade}, {Bowman}, {Granzer}, {Kallinger}, {Koudelka}, {Lovekin}, {Neiner}, {Pablo}, {Pigulski}, {Popowicz}, {Ramiaramanantsoa}, {Rucinski}, {Strassmeier}, \& {Wade}}]{Weiss_2021}
{Weiss}, W.~W., {Zwintz}, K., {Kuschnig}, R., {et~al.} 2021, \bibinfo{title}{{Space Photometry with BRITE-Constellation},} Universe, 7, 199, \dodoi{10.3390/universe7060199}

\bibitem[{G.~N. Whittaker(2014)Whittaker}]{Whittaker_2014}
Whittaker, G.~N. 2014, PhD thesis, University of Birmingham

\bibitem[{J.~N. {Winn} {et~al.}(2011){Winn}, {Matthews}, {Dawson}, {Fabrycky}, {Holman}, {Kallinger}, {Kuschnig}, {Sasselov}, {Dragomir}, {Guenther}, {Moffat}, {Rowe}, {Rucinski}, \& {Weiss}}]{Winn_2011}
{Winn}, J.~N., {Matthews}, J.~M., {Dawson}, R.~I., {et~al.} 2011, \bibinfo{title}{{A Super-Earth Transiting a Naked-eye Star},} \apjl, 737, L18, \dodoi{10.1088/2041-8205/737/1/L18}

\bibitem[{K.~T. {Wraight} {et~al.}(2011){Wraight}, {White}, {Bewsher}, \& {Norton}}]{Wraight_2011}
{Wraight}, K.~T., {White}, G.~J., {Bewsher}, D., \& {Norton}, A.~J. 2011, \bibinfo{title}{{STEREO observations of stars and the search for exoplanets},} \mnras, 416, 2477, \dodoi{10.1111/j.1365-2966.2011.18599.x}

\bibitem[{L.-C. {Yeh} \& I.-G. {Jiang}(2020){Yeh} \& {Jiang}}]{Yeh_2020}
{Yeh}, L.-C., \& {Jiang}, I.-G. 2020, \bibinfo{title}{{Searching for Possible Exoplanet Transits from BRITE Data through a Machine Learning Technique},} \pasp, 133, 014401, \dodoi{10.1088/1538-3873/abbb24}

\bibitem[{K. {Yumoto} {et~al.}(2024){Yumoto}, {Tatsumi}, {Kouyama}, {Golish}, {Cho}, {Morota}, {Kameda}, {Sato}, {Rizk}, {DellaGiustina}, {Yokota}, {Suzuki}, {de Le{\'o}n}, {Campins}, {Licandro}, {Popescu}, {Rizos}, {Honda}, {Yamada}, {Sakatani}, {Honda}, {Matsuoka}, {Hayakawa}, {Sawada}, {Ogawa}, {Yamamoto}, {Lauretta}, \& {Sugita}}]{Yumoto_2024}
{Yumoto}, K., {Tatsumi}, E., {Kouyama}, T., {et~al.} 2024, \bibinfo{title}{{Comparison of optical spectra between asteroids Ryugu and Bennu: I. Cross calibration between Hayabusa2/ONC-T and OSIRIS-REx/MapCam},} \icarus, 417, 116122, \dodoi{10.1016/j.icarus.2024.116122}

\end{thebibliography}
\bibliographystyle{aasjournalv7}



\end{document}